\documentclass[preprint]{aastex}
\usepackage{amsmath}
\usepackage{natbib}

\newcommand{\mppc}[1]{\:M_\odot {\rm pc}{}^{-#1}}
\def\del{\partial}
\def\obs{{\rm obs}}

\begin{document}
\title{Updated Kinematic Constraints on a Dark Disk}
\author{Eric David Kramer and Lisa Randall}
\affil{Department of Physics, Harvard University, Cambridge, MA, 02138}
\email{ekramer@physics.harvard.edu}
\email{randall@physics.harvard.edu}

\begin{abstract}
We update the method of the \citet{hf2000} study, including an updated model of the Milky Way's interstellar gas, radial velocities, an updated reddening map, and a careful statistical analysis, to bound the allowed surface density and scale height of a dark disk. We  pay careful attention to the self-consistency of the model, including the gravitational influence of the dark disk on other disk components, and to the net velocity of the tracer stars. We find that the data set exhibits a non-zero bulk velocity in the vertical direction as well as a displacement from the expected location at the Galactic midplane. If not properly accounted for, these features would bias the bound toward low dark disk mass. We therefore perform our analysis two ways. In the first, traditional method, we subtract the mean velocity and displacement from the tracers' phase space distributions. In the second method, we perform a non-equilibrium version of the HF method to derive a bound on the dark disk parameters for an oscillating tracer distribution. Despite updates in the mass model and reddening map, the traditional method results remain consistent with those of HF2000. The second, non-equilibrium technique, however, allows a surface density as large as  $14 \mppc{2}$ (and as small as $0\,\mppc{2}$), demonstrating much weaker constraints. For both techniques, the bound on surface density is weaker for larger scale height. In future analyses of Gaia data, it will be important to verify whether the tracer populations are in equilibrium.
\end{abstract}

\keywords{Galaxy: kinematics, disk, solar neighborhood; cosmology: dark matter}

\section{Introduction}
\label{sec:intro}

Since the original study by \citet{oort1,oort2}, the question of disk dark matter has been a subject of controversy. Over the years, several authors have suggested the idea of a dark disk to explain various phenomena. \citet{kalb07} proposed a thick dark disk as a way to explain the flaring of the interstellar gas layer. \citet{dd1} showed using cosmological simulations that a thick dark disk is formed naturally in a $\Lambda$CDM cosmology as a consequence of satellite mergers. A phenomenologically very different idea is that of a thin dark disk. \citet{dddm} put forward a model for dark matter, coined Double Disk Dark Matter (DDDM), where a small fraction of the total dark matter is self-interacting and dissipative, forming a thin disk. Following the 2013 DDDM paper, \citet{dino} showed that a dark matter disk of surface density $\sim$10 $\mppc{2}$ and scale height $\sim 10$ pc could potentially explain periodicity of comet impacts on earth, giving a target surface density and scale height. If such a dark disk were found to exist, we would want to know what are its surface mass density $\Sigma_D$ and its scale height $h_D$. Its density in the plane would then be approximately  $\rho_D(0) \simeq \Sigma_D/4h_D$.\footnote{Note that in our convention for scale height $h$, a self-gravitating isothermal disk \citep{spitzer} is defined to have the spatial dependence $\rho(z) \sim {\rm sech}^2(z/2h)$. At large $z$, this is proportional to $e^{-|z|/h}$.} A dark disk model allows us to test the viability of these  benchmark values. We do not assume this model is favored, but since the literature clearly supports no dark disk as a possibility, we ask whether it also allows for a disk with this or greater density.

A possible concern in a thin dark-disk model is if the system can emerge dynamically when instabilities and fragmentation are accounted for. These concerns may be resolved with simulations and more careful theoretical considerations, or possibly additional model building. For the purposes of this paper, however, we consider the dark disk from a purely phenomenological perspective: we ask only what the data tell us about the presence of a thin dark disk, and how to hope to better determine this in the future.

The existing literature suggests that such a model  is  highly constrained. In principle, there are many ways in which one might aim to constrain a dark disk kinematically. All of these rely on the Poisson-Jeans theory (cf. Section \ref{sec:theory}). Based on our survey of the literature, we categorize the attempted constraints into three main categories:

\begin{enumerate}
\item Using stellar kinematics to fit a model for known matter containing interstellar gas, Galactic disk, and dark halo. The parameter being fit here is the surface density of the Galactic disk. Once this is found, one has a measure of the total surface density of the galactic disk.  The authors then compare this result to the inventories of known matter to try to constrain the dark disk surface density.

 \item Using stellar kinematics to directly measure the vertical gravitational acceleration $K_z$ as a function of height $z$ above the Galactic plane. By the Poisson equation, this is proportional to the total surface density $\Sigma_z$ integrated to height $z$ above the plane. Again an attempt is made to compare this result to the  inventories of known matter.

\item Using an isothermal component model \citep{bahc84c,bahc84b} for known matter and fitting the remaining dark matter mass self-consistently using stellar kinematics.
\end{enumerate}

Table \ref{tab:constraints} contains a list of such constraints and which category they fall into. In the face of all these constraints, one may wonder what hope is left for a dark disk. We claim, however, that, at least in their present state, the majority if not all of the constraints in Table \ref{tab:constraints} do not apply.

Method 1 assumes that there is no dark disk and fits a model using known matter to stellar kinematics to show that a dark disk is not necessary. For example, \citet{KG89b,KG89a,KG91} showed that the density and velocity distributions of K-dwarfs in the Milky Way were consistent with  little or no disk dark matter (besides the usual dark halo). Although it is correct that the stellar kinematics do not require a dark disk, the converse is not correct: they do not exclude a dark disk. One might then try to exclude a dark disk by comparing the total Galactic disk surface density found in this way to the inventories of known matter, which at face value seems to be correct. For example, using the distribution function techniques developed by Binney \citep{bt}, \citet{bovy14} dynamically constrained the amount of matter in the Galactic disk, claiming that their results, which are consistent with only standard Galactic components, ``leave little room for a dark disk component" in the Milky Way. 

\begin{figure}
\plotone{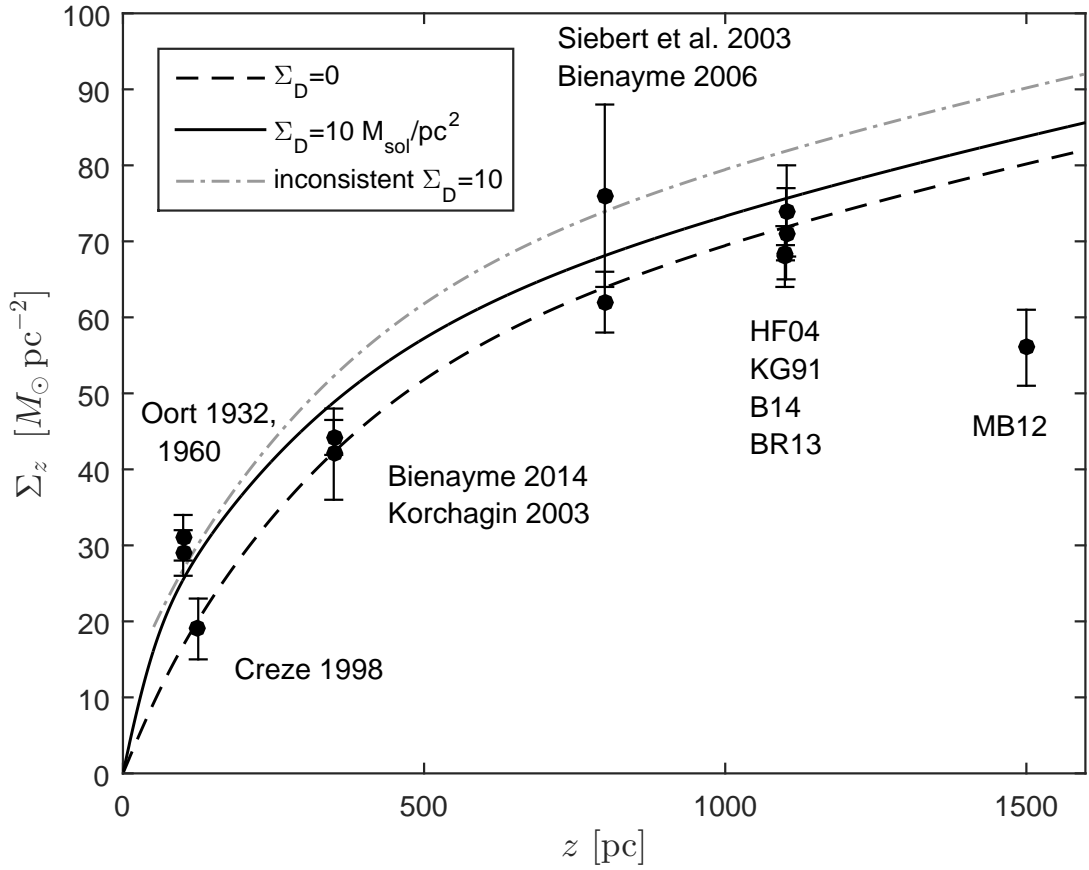}
\centering{}\caption{{ $\Sigma_{\rm tot}(z)$ curves for models with $\Sigma_D=0$ (dashed line) and $\Sigma_D=10\mppc{2}$ (solid line). Curve for inconsistent inclusion of dark disk without pinching back-reaction also shown (grey, dot-dashed line).
}}
\label{fig:naive}
\end{figure}

However, there is an important detail that has so far been overlooked: the dark disk can actually make room for itself. This is because the distributions of known components in the Galactic disk are extrapolated from observations near the midplane. However, depending on the gravitational pull in the disk, this extrapolation can produce thinner scale heights for these components. For example, suppose we know the density $\rho_i$ of certain mass components near the plane. The surface density contributions of these components will be $\Sigma_i \sim \rho_i h_i$, where $h_i$ is the thickness of components $i$. The dark disk will affect this value by `pinching' the matter distributions and reducing the thickness $h_i$ of the components, resulting in a lower surface density determination. The surface density estimation for known matter is therefore lower in the presence of a dark disk than without one. Qualitatively, we can write this `triangle inequality' as
\begin{eqnarray}
\label{eq:triangle1}
\Sigma\,(\textrm{ visible matter + dark disk })\, \leq \,\Sigma\,(\textrm{ visible matter alone }) + \Sigma\,(\textrm{ dark disk }).
\end{eqnarray}
We therefore need to keep this in mind when comparing kinematic determinations of the total surface density to known mass inventories. Figure \ref{fig:naive} compares the various bounds in the literature to models with and without a dark disk, as well as to a dark disk model that does not take into account the pinching effect back-reaction of the dark disk.

An additional problem with this method is that the specific models that are fit to the kinematics do not allow for a dark disk. Most of the studies using Method 1 \citep{KG89b,KG89a,KG91,bienayme06,zhang,bienayme14} fit models containing only 	 and halo, but no thin component. Other studies \citep{zhang,bovy14} considered a thin gas component as well but kept its mass fixed and did not allow for any additional mass in a thin component. The study of \citet{creze}, in fact, did not consider a stellar disk component at all and assumed a constant density potential. Their result is therefore very difficult to compare to the literature and in particular to a thin dark disk model.

A similar argument would apply to Method 2. Indeed, the results of \citet{bovytr} are consistent with a thin dark disk scenario for precisely the same reason. However, it should be noted that the results of \citeauthor{bovytr} were based on kinematic data high above the Galactic midplane (1.5-4.5 kpc). Uncertainties in asymmetric drift at these heights could possibly allow total surface densities that are much lower. In fact, analyzing the same data, \citet{mb12} find that the data do not even allow a dark halo, let alone a dark disk. In \citet{mb15}, however, thery point out that this method is effective only in constraining the mass above $z=1.5$ kpc, i.e. $\Sigma_{\rm tot}(z)-\Sigma_{\rm tot}({\rm 1.5\,kpc})$. (This must be the case, as their 2012 value of $\Sigma_{\rm tot}({\rm 1.5\,kpc})$ is significantly lower than all literature measurements of $\Sigma_{\rm tot}({\rm 1.1\,kpc})$.) This argument applies to the results of \citet{bt} as well. These results therefore do not apply in constraining a thin dark disk. 

Another study measuring $\Sigma_{\rm tot}(z)$ directly is that of \citet{korchagin}. Their result $\Sigma_{\rm tot}({\rm 10 \, pc})=10\pm1\mppc{2}$ is highly dependent on their choice for the form of $\rho_{\rm tot}(z)$. By assuming that $\rho_{\rm tot}(z)\sim {\rm sech}(z/2h)$ or ${\rm sech}^2(z/2h)$, they are actually not allowing a thin massive component from the outset. Their only robust result is therefore $\Sigma_{\rm tot}({\rm 350\,pc})=42\pm6\mppc{2}$ since it does not depend on the type of extrapolation to $z=0$ used. Even at 350 pc, these results seem to be in disagreement with a dark disk of the size we are considering, as can be seen in Figure \ref{fig:naive}. However, it was pointed out to us (T. Girard, private communication) that the extinction corrections were applied with the wrong sign. Intuitively, without reddening corrections, we expect preferential reddening near the Galactic plane to increase the apparent number of red giants near the plane, giving a cuspy profile such as we would expect as arising in the presence of a thin dark disk. Although reddening corrections should remove this potentially large bias, applying the reddening corrections with the wrong sign should make it twice as large. It is therefore unclear why \citeauthor{korchagin} did not obtain a tighter bound on $\Sigma_{\rm tot}(50\,{\rm pc})$. At any rate, we cannot take their present results at face value.

On the other hand, the older studies of \citet{oort1,oort2} seem to particularly favor a dark disk model, as can be seen in Figure \ref{fig:naive}. (Here, we assume the robust result to be $\Sigma_{\rm tot}({\rm 100\,pc})$ and ignore the endpoint result $\Sigma_{\rm tot}({\rm 50\,pc})$.)  However, \citet{readrev} argues that these results were based on `poorly calibrated photometric distances', stars that were too young to be in equilibrium, and other questionable assumptions.

An important lesson here is that, as with particle physics experimental studies, it helps to have a definite model in order to find a self-consistent bound. Previous analyses may not apply because of changes in measured parameters, but also because the constraint itself depends on the model, which determines the distribution of other components. On top of this, most previous studies didn't use the dark disk thickness as an independent parameter. With a model, it becomes clear that it makes sense to analyze the data with such a parameter included. Given the large amount of data becoming available, it makes sense to view the data as a way of measuring standard parameters, but also constraining--or hopefully discovering--new ones. It is also true that, were one to analyze a different model aside from the dark disk model we have in mind, one may have to do the analysis differently to account for any different parameters that we (and previous authors) did not include.

Method 3, on the other hand, can include a dark disk in a manner that is gravitationally self-consistent. For example, \citet{bahc84c} used this method to show that a thin dark disk similar in size to the interstellar gas disk had to be lighter than $\Sigma_D=17\mppc{2}$ in the solar region. \citet{brc} also used this technique to show that a thick dark disk had to be lighter than $\rho_{D}(0)\leq0.03\mppc{3}$, as did \citet{ff} to show that a model with no dark halo but a very light thick dark disk, with a total glactic surface density $\Sigma_{\rm tot}=52\pm13$, was a good match to the kinematics of K giants in the solar region. Most importantly, this method was used by Holmberg \& Flynn (2000) to show that the kinematics of A and F stars in the solar region were consistent with visible matter distributions alone, without the need for a dark disk. By adding dark matter to known components, they were able to compute a midplane density for the total matter of $\rho(0)=0.100 \pm 0.006 \mppc{3}$. Since they considered dark matter as thin as the molecular hydrogen mass component, their result effectively sets a bound on a dark disk of scale height of 40 pc. 

We will see that, using Method 3, the bounds on a dark disk become even tighter as the scale height is decreased. However, there are at least two questions to be asked on the interpretation of this result. i) Can the combined errors in visible components and kinematics allow for a dark disk on their own? ii) Do the assumptions of their kinematic methods (e.g. of thermal equilibrium) hold?

\begin{table}
\centering{}\caption{Bounds on the surface density and local density of total matter and visible matter in the galaxy. }
\label{tab:constraints}
\begin{tabular}{lclc}
\tableline\tableline
Authors & Year & \:\:\:Bound $\;\; [\!\mppc{n}]$ & Category\\ 
\tableline\tableline
\citeauthor{oort1}	& \citeyear{oort1} & $\Sigma_{\rm tot}({\rm 100\,pc})$	= 31& 2\\ 
\citeauthor{oort2}	& \citeyear{oort2} & $\Sigma_{\rm tot}({\rm 100\,pc})=29\pm10\%$ & 2\\
\citeauthor{bahc84c}	& \citeyear{bahc84c} & $\Sigma_{D,{\rm thin}}\leq17\qquad\rho_{\rm tot}(0)\leq0.24$& 3\\
\citeauthor{bahc84b}	& \citeyear{bahc84b} & $\Sigma_{\rm tot}=55-83\qquad\rho_{\rm tot}(0)=0.17-0.25$ & 3\\ 
\citeauthor{brc}	& \citeyear{brc}	& $\rho_{\rm DM}(0)\leq0.03\;\;$   for thick dark disk & 3 \\ 
\citeauthor{KG91}	& \citeyear{KG91} & $\Sigma_{\rm tot}({\rm 1.1\, kpc})=71\pm6 $ & 1\\
\citeauthor{bahc92}	& \citeyear{bahc92} & $\Sigma_{\rm tot} = 70^{+24}_{-16}$ & 3 \\ 
\citeauthor{ff}	& \citeyear{ff} & $\Sigma_{\rm tot}=52\pm13$ & 3* \\
\citeauthor{pham} & \citeyear{pham} & $\rho_{\rm tot}(0)=0.11\pm0.01$ & NA \\ 
\citeauthor{creze} & \citeyear{creze}& $\rho_{\rm tot}(0)=0.076\pm0.015$ (assumed constant density) & 1* \\ 
\citeauthor{hf2000}	& \citeyear{hf2000} & $\rho_{\rm tot}(0)=0.102\pm0.010\qquad\rho_{\rm vis}=0.095$ & 3* \\ \citeauthor{korchagin}	& \citeyear{korchagin} & $\Sigma_{\rm tot}({\rm 350\, pc})=42\pm6$ & 2 \\ 
\citeauthor{siebert}& \citeyear{siebert} & $\Sigma_{\rm tot}({\rm 800\, pc})=76^{+25}_{-12}$ & 1\\ 
\citeauthor{hf2004}	& \citeyear{hf2004} & $\Sigma_{\rm tot}({\rm 1.1\, kpc})=74\pm6$& 3\\
\citeauthor{bienayme06}	& \citeyear{bienayme06} & $\Sigma_{\rm tot}({\rm 800\, pc})=57-66$ & 1\\
\citeauthor{grl11}	& \citeyear{grl11} & $\rho_{\rm halo}=0.003-0.033$ & 3\\
\citeauthor{mb12b}& \citeyear{mb12b} & $\Sigma_{\rm tot}({\rm 1.5\,kpc})=55.6\pm4.7$ & 2 \\
\citeauthor{bovytr}& \citeyear{bovytr} & $\rho_{\rm halo}=0.008\pm0.003$ & 2 \\
\citeauthor{zhang}& \citeyear{zhang} & $\Sigma_{\rm tot}({\rm 1\,kpc})=67\pm6 $ \\
&  & $\rho_{\rm halo}(0)=0.0065\pm0.0023$  & 1\\
\citeauthor{bovy14} & \citeyear{bovy14} & $\Sigma_{1100}=68\pm4$ & 1 \\
\citeauthor{bienayme14} & \citeyear{bienayme14} & $\Sigma_{\rm tot}({\rm 1.1\, kpc})=68.5\pm1$\\
& & $\Sigma_{\rm tot}({\rm 350\, pc})=44.2^{+2.3}_{-2.9}$ & 1\\
\tableline
&&* denotes bounds derived using HF technique
\end{tabular}
\end{table}

To answer the first question we note that a non-zero dark disk mass could certainly have been hiding in the previously assumed error bars on $\Sigma_{\rm vis}$, including the errors on the interstellar gas, which had until now been around 50\%. \citet{bovy14} reported an uncertainty on $\Sigma_{\rm tot}$ of $\pm4\mppc{2}$, and an uncertainty on $\Sigma_{\rm stars\;+\;remnants}$ also of $\pm4\mppc{2}$. If we combine this with the traditionally assumed $\pm7\mppc{2}$ error on the interstellar gas, we can already allow a dark disk with mass $\sqrt{4^2+4^2+7^2}\simeq9\mppc{2}$. 

The baryonic matter components of the Galaxy have, however, been more carefully measured in recent years. Revised values for the interstellar gas parameters are the subject of \citet{paper2}. We also include new values for the stellar components, revised by \citet{mckee}. We repeated the study of HF2000 including radial velocites (which were not available at the time) and using the recent three-dimensional reddening map of \citet{schlafly}. With these updates, we show in this paper that the traditional HF2000 method would not allow a dark disk heavier than $\sim4\mppc{2}$.

Regarding the second question, however, we note that one of the major assumptions in the kinematic method of HF2000 is that the populations under study are in statistical-mechanical equilibrium. We claim (and show in Appendix \ref{sec:noneq}) that the definition of equilibrium needed for these analyses is one that precludes the possibility of oscillating solutions, in which the tracer population is in a distribution whose center oscillates above and below the Galactic plane. Such behavior, if present, would smooth out the effects of any thin Galactic component, in which case incorrectly assuming a static distribution would  result in an upper bound on a dark disk that is too low. 

We show that the tracer populations studied by HF2000 indeed do possess a net vertical velocity as well as a center that is vertically displaced from the Galactic midplane, which are evidence for such deviations from equilibrium. The HF2000 analysis, which assumes that the tracer populations are in equilibrium, produces an overly strong bound. We show here how a consistent analysis can be performed without any need for the assumption of equilibrium. Analyzed in this way, we show that the kinematics currently allow for a thin dark disk of up to $14 \mppc{2}$, with a weaker bound for thicker disks. (These considerations apply equally to the studies of \citet{ff} and of \citet{creze}. The latter sample, in fact, does show deviations from equilibrium including a negative net vertical displacement from the midplane, as reported by the authors, and a net vertical velocity, as can be inferred from their numbers. This may, along with the reasons described above, account for the unusually low values of $\rho_{\rm tot}(0)$ found by these authors.)

\section{Theory}
\label{sec:theory}

The Poisson-Jeans theory will be important both for constructing a model of the Galactic potential and for gaining a qualitative understanding of the effect of a thin dark disk. We will now explain the Poisson-Jeans theory, following which we will use the P-J theory to solve a simple toy model. We will then explain the theory behind the HF2000 study, and derive the Holmberg \& Flynn relation.

\subsection{Poisson-Jeans Theory}
Consider the phase-space distribution $f_i({\bf x},{\bf v})$ for a stellar population $i$, satisfying
\begin{eqnarray}
\int\! d^3v\, f_i({\bf x},{\bf v})= \rho_i({\bf x})
\end{eqnarray}
and, for a more general function $g$,
\begin{eqnarray}
\int\! d^3v\, f_i({\bf x},{\bf v})g({\bf v})= \rho_i({\bf x})\,\langle g\rangle
\end{eqnarray}
where $\rho_i({\bf x})$ is the population density. The $f_i$ must also satisfy the collisionless Boltzmann equation (Liouville's theorem)
\begin{equation}
\label{eq:boltzmann}
\frac{D f_i}{Dt} \equiv \frac{\del f_i}{\del t} + \dot{{\bf x}}\cdot\frac{\del f_i}{\del {\bf x}} + \dot{{\bf v}}\cdot\frac{\del f_i}{\del {\bf v}}=0.
\end{equation}
If we assume our stellar population is in equilibrium then the first term on the right vanishes. Also, we can replace $\dot{{\bf x}}$ and $\dot{{\bf v}}$ by ${\bf v}$ and $-{\del \Phi}/{\del {\bf x}}$ respectively, where $\Phi$ is the gravitational potential. From the Boltzmann equation, we can derive the vertical Jeans equation in the standard way, found e.g. in \citet{bt}. The vertical Jeans equation then reads:
\begin{eqnarray}
\label{eq:tilt}
\frac{1}{R}\frac{\del}{\del R}(R\rho_i \sigma_{i,Rz}) + \frac{\del}{\del z}(\rho_i \sigma_{i,z}^2) + \rho_i \frac{\del \Phi}{\del z} = 0
\end{eqnarray}
The first term in Equation \ref{eq:tilt} describes the rate of change of the correlation between $v_z$ and $v_R$, so it is commonly referred to as the `tilt' term. The tilt term is expected to be very small near the plane, since, being asymmetric in $z$, it must vanish at $z=0$ \citep{mb12b}. If we restrict our analysis to heights that are small compared to the tracer population's scale height, we can safely neglect this term. We thus have (dropping the subscript on $\sigma_{i,z}$)
\begin{eqnarray}
\frac{\del}{\del z}(\rho_i \sigma_i^2) + \rho_i \frac{\del \Phi}{\del z} = 0
\end{eqnarray}
which we can rewrite as
\begin{eqnarray}
\frac{1}{\rho_i \sigma_i^2}\frac{\del}{\del z}(\rho_i \sigma_i^2) + \frac{1}{\sigma_i^2} \frac{\del \Phi}{\del z} = 0,
\end{eqnarray}
from which we can immediately see that the solution is
\begin{eqnarray}
\rho_i\, \sigma_i^2 \propto \exp\left(-\int\! dz\, \frac{1}{\sigma_i^2}\frac{\del \Phi}{\del z}\right).
\end{eqnarray}
If we additionally take our stellar population to be isothermal, that is $\sigma_i(z) = {\rm constant}$, then the solution reduces further to
\begin{eqnarray}
\rho_i(R,z)=\rho_i(R,0)\,e^{-{\left(\Phi(R,z)-\Phi(R,0)\right)}/{\sigma_i^2}}.
\end{eqnarray}
We will now consider only $R=R_\odot$, where we will fix $\Phi(R_\odot,0)=0$. This gives
\begin{eqnarray}
\label{eq:ideal}
\rho_i(z)=\rho_i(0)\,e^{-{\Phi(z)}/{\sigma_i^2}}.
\end{eqnarray}
Indeed, once we have assumed equilibrium, vanishing tilt, and isothermality, what we are left with is simply a gas at temperature per unit mass $  kT_i/M = \sigma_i^2$, and Equation \ref{eq:ideal} is merely a Boltzmann factor.
A possible objection to the method is that it was found by \citet{grl11} that neglecting the tilt term in Equation \ref{eq:tilt} in general leads to a biased determination of $\rho_{\rm dm}$. However, this was only found to be a problem at heights of $z\gtrsim 0.5\, {\rm  kpc}$ or greater. \citeauthor{grl11} describe the HF2000 sample, which is relatively close to the plane, as `unlikely to be biased'. With this solution to the Jeans equation, we can now relate the potential $\Phi$ to the mass distribution consisting of the different components $\rho_i$. By the Poisson equation,
\begin{eqnarray}
\nabla^2 \Phi = 4\pi G \sum_i \rho_i.
\end{eqnarray}
We can split up the Laplacian in the standard way, following \citet{KG89a}:
\begin{mathletters}
\begin{eqnarray}
\label{eq:poisson1}
\nabla^2 \Phi&=&\frac{\del^2\Phi}{\del z^2}+\frac{1}{R}\frac{\del}{\del R}\left(R\frac{\del \Phi}{\del R}\right)\\
&=&\frac{\del^2\Phi}{\del z^2}+\frac{1}{R}\frac{\del}{\del R}V_c^2\\
&=&\frac{\del^2\Phi}{\del z^2}+2\frac{V_c}{R}\frac{\del V_c}{\del R}\\
&=&\frac{\del^2\Phi}{\del z^2} + 2(B^2-A^2)
\end{eqnarray}
\end{mathletters}
where $V_c$ is the local circular velocity and $A$, $B$ are the `Oort constants'
%\notetoeditor{The reviewer feels that $V_c^2=-R F_R$ should only be valid in the plane, as pointed out by \citet{mb14} regarding \citet{bovytr}. We note, however, that we are using this as our definition of $V_c$ both in and away from the plane, and that this notation seems to be standard in the literature. Is the reviewer referring to an alternate definition of $V_c$?} 
\begin{eqnarray}
A\equiv \frac12\left(-\frac{\del V_c}{\del R} + \frac{V_c}{R}\right)\\
B\equiv -\frac12\left(\frac{\del V_c}{\del R} + \frac{V_c}{R}\right).
\end{eqnarray}
We have assumed azimuthal symmetry in Equation \ref{eq:poisson1}. Although in reality the azimuthal symmetry of the Galaxy is spoiled by spiral arms and other structures, we will assume that these are {not relevant} over the time scales needed for our tracer populations to relax to their current distributions. At the Sun's position, the second term in Equation \ref{eq:poisson1} is very small,  so we shall rename it $4\pi G \delta\rho$. From \citet{catul}, we have $A-B=29.45\pm0.15 \,{\rm km \,s^{-1} kpc^{-1}}$ and $A+B=0.18 \pm 0.47\,{\rm km \,s^{-1} kpc^{-1}}$, giving
\begin{eqnarray}
\delta\rho = (-2\pm5)\times10^{-4} M_\odot {\rm pc^{-3}}
\end{eqnarray}
near the $z=0$ plane. This is comparable in magnitude to the density of red giant stars and to the stellar halo density \citep{hf2000}, but is about 2-3 orders of magnitude smaller than the total density, and can safely be neglected:
\begin{eqnarray}
\label{eq:poisson2}
\frac{\del^2\Phi}{\del z^2} = 4\pi G \sum_i \rho_i.
\end{eqnarray}
Combining Equation \ref{eq:poisson2} with the Jeans equation (\ref{eq:ideal}), we have the Poisson-Jeans equation for the potential $\Phi$:
\begin{eqnarray}
\label{eq:poissonjeans1}
\frac{\del^2\Phi}{\del z^2} = 4\pi G \sum_i \rho_i(0) e^{-\Phi/\sigma_i^2}.
\end{eqnarray}
where we have dropped the $R$ coordinate label. This can also be cast in integral form (assuming $z$-reflection symmetry)
\begin{eqnarray}
\label{eq:poissonjeans2}
\frac{\rho_i(z)}{\rho_i(0)} = \exp\left(-\frac{4\pi G}{\sigma_i^2}\sum_k \int_0^{z}\!dz^{\prime}\,\int_0^{z^{\prime}}\!dz^{\prime\prime}\, \rho_k(z^{\prime\prime})\right)
\end{eqnarray}
which is the form used in our Poisson-Jeans solver.

\subsection{A Toy Model}
\label{sec:toy}
We now study the Poisson-Jeans theory in some simple models following \citet{spitzer}. This will be useful in understand the qualitative effects of a thin dark disk. We first study a self-gravitating, single-component galaxy. In this case, Equation \ref{eq:poissonjeans1} reads
\begin{eqnarray}
\frac{1}{4\pi G} \frac{d^2 \Phi}{d z^2} = \rho_0 e^{-\Phi(z)/\sigma^2}
\end{eqnarray}
which has solution
\begin{eqnarray}
\frac{d\Phi(z)}{dz}=\frac{\sigma^2}{h} \tanh(z/2h)
\end{eqnarray}
or
\begin{eqnarray}
\label{eq:spitzersol}
\rho(z)=\rho_0\, {\rm sech}^2(z/2h)
\end{eqnarray}
where
\begin{eqnarray}
\label{eq:h}
h=\frac{\sigma}{\sqrt{8\pi G \rho_0}}\;.
\end{eqnarray}
Although this was a toy model, it illustrates two very important features of gravitating disks: 1) The scale height $h$ of the disk grows with its vertical dispersion $\sigma$, approximately as $h\sim \sigma$, and 2) The scale height decreases with the total midplane density. These features will remain qualitatively true even when more components are included.

\begin{figure}[h]
\plotone{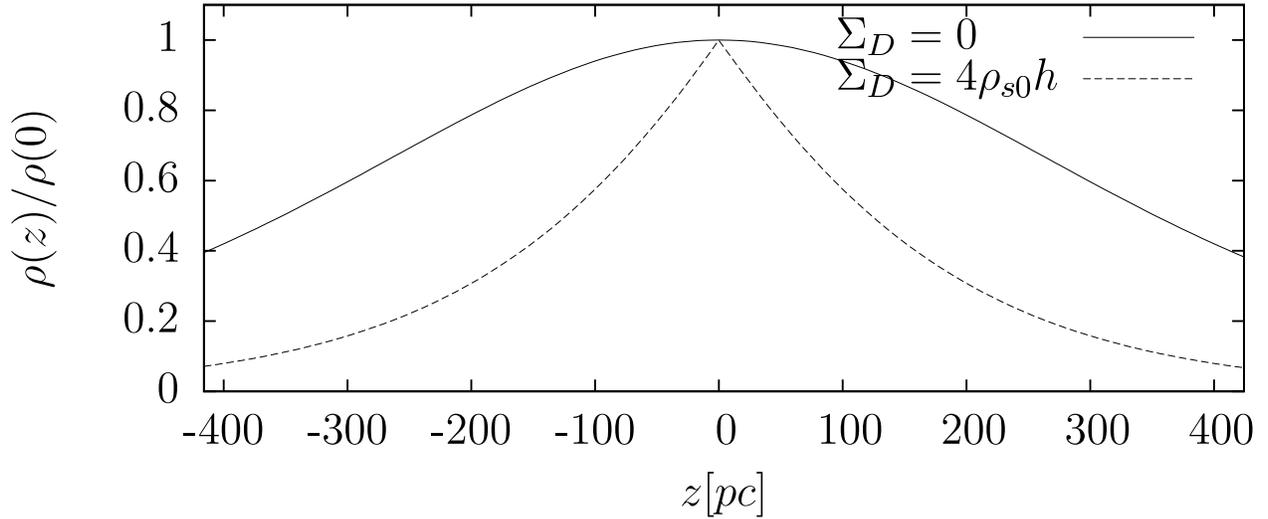}
\centering{}\caption{{ A plot of the exact solutions without and with a dark disk of $Q=1$. The density is `pinched' by the disk, in accordance with Equation \ref{eq:exact}.}}
\label{fig:toy}
\end{figure}

We can now ask what happens if we include an additional very thin, delta-function component (as a toy approximation to a dark disk). In this case, we have
\begin{eqnarray}
\nabla^2 \Phi &=& 4\pi G \rho_s + 4\pi G \Sigma_D \delta(z)
\end{eqnarray}
and the Poisson-Jeans equation takes the form
\begin{eqnarray}
\label{eq:pj2cmpt}
\frac{1}{4\pi G}\frac{d^2\Phi(z)}{dz^2}-\Sigma_D\;\delta(z) =\rho_{s0}\exp\left(-\frac{\Phi(z)}{\sigma^2}\right)
\end{eqnarray}
where $\rho_s$ stands for the density of stars and $\Sigma_D$ is the dark disk surface density. Defining $Q\equiv \Sigma_D/4\rho_{s0}h_s$, (with $h_s$ defined as in Equation \ref{eq:h}), we can write down the exact solution, which is
\begin{eqnarray}
\label{eq:exact}
\rho_s(z)=\rho_{s0} (1+Q^2)\; {\rm sech}^2\!\left( \frac{\sqrt{1+Q^2}}{2h_s}\left(|z| + z_0\right)\right)
\end{eqnarray}
with
\begin{eqnarray}
z_0 \equiv \frac{2h_s}{\sqrt{1+Q^2}} {\rm arctanh}\left(\frac{Q}{\sqrt{1+Q^2}}\right).
\end{eqnarray}
Effectively, the dark disk has the effect of `pinching' the density distribution of the other components, as we can see in Figure \ref{fig:toy}. In particular, it reduces their scale heights, and for a fixed midplane density $\rho_i(0)$, it implies that their surface density $\Sigma_i$ is less than what it would be without the disk. Actually, this is what would happen were we to include any other additional  mass component.

An important consequence of this for the stellar disk, especially in the context of our analysis, is that for a fixed midplane density, the effect of the dark disk is to $decrease$ the stellar surface density. Integrating Equation \ref{eq:exact}, the stellar surface density find is
\begin{mathletters}
\begin{eqnarray}
\Sigma_s(Q)	&=&\Sigma_s(0)\left(\sqrt{1+Q^2}-Q\right)\\
		&=&\sqrt{\Sigma_s(0)^2+\Sigma_D^2\;}\,-\,\Sigma_D
\end{eqnarray}
\end{mathletters}
where $\Sigma_s(0)\equiv 4\rho_{s0}h_s$ is what the surface density would have been without the dark disk. We can see that the stellar surface density, $\Sigma_s(Q)$ is monotonically decreasing with $Q$. We can then write the total surface density as
\begin{mathletters}
\begin{eqnarray}
\Sigma_{\rm tot}&=&\Sigma_s(Q)+\Sigma_D\\
		&=&\sqrt{\Sigma_s(0)^2+\Sigma_D^2}
\label{eq:triangle2}
\end{eqnarray}
\end{mathletters}
verifying the triangle inequality (Equation \ref{eq:triangle1}) for this simple case.

\subsection{The HF study}
\label{sec:hf2000}
\begin{figure}
\plotone{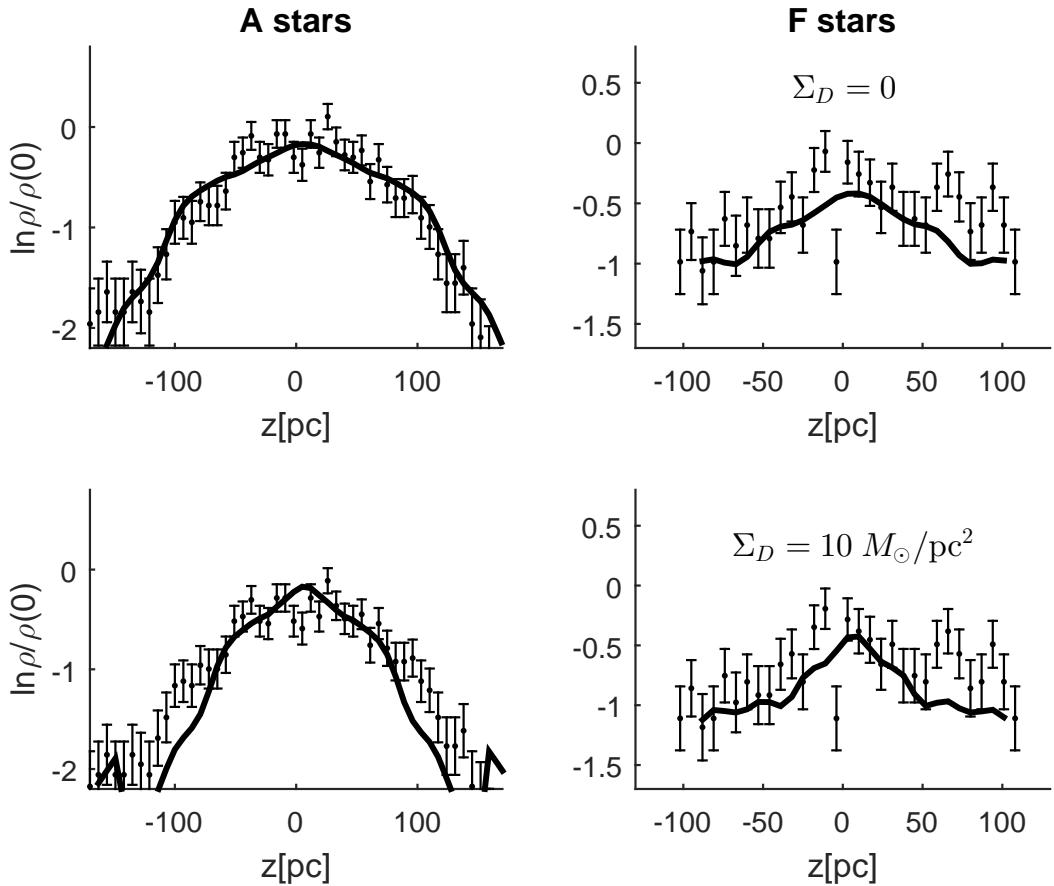}
\centering{}\caption{{ $Top$: The HF2000 study. The HF2000 model with no disk dark matter agrees quite well with the A and F star data. $Bottom$: The HF2000 result, this time including a dark disk with $\Sigma_D=10\mppc{2}$ and $h_D=10$ pc.
}}
\label{fig:hf2000}
\end{figure}

We  now explain the analysis of \citet{hf2000}, and derive the HF relation. This analysis, developed under different forms by \citet{KG89b,KG89a}, \citet{fw}, \citet{ff}, \citet{hf2000}, is related to the P-J theory. 
Near $R=R_\odot$, we can write
\begin{eqnarray}
f({\bf x},{\bf v})\propto f_z(z,v_z).
\end{eqnarray}
where $f(z,v_z)$ represents the one-dimensional phase space density in $z,v_z$. We therefore work solely with $f_z(z,v_z)$ and drop the subscript $z$. This function satisfies
\begin{eqnarray}
\int \!dv_z\, f(z,v_z) \, g(v_z) = \rho(z) \langle\, g \,\rangle.
\end{eqnarray}
The Boltzmann equation also tells us that
\begin{eqnarray}
\label{eq:boltz}
v_z\frac{\del f}{\del z} - \frac{\del \Phi}{\del z}\frac{\del f}{\del v_z} = 0
\end{eqnarray}
which has solution
\begin{eqnarray}
\label{eq:ez}
f(z,v_z)=F\left(\frac12 v_z^2 + \Phi(z) \right).
\end{eqnarray}
That is, since this function is constant it must be a function solely of the `vertical energy' $\frac12 v_z ^2 + \Phi(z)$. Using Eqs. \ref{eq:boltz} and \ref{eq:ez} we can write
\begin{mathletters}
\begin{eqnarray}
\rho_i(z) &=& \int_{-\infty}^{\infty} \! dw \, f_i(z, w)\\
	&=& \int_{-\infty}^{\infty}\! dw \, f_i(0,\sqrt{w^2 + 2\Phi(z)})
\end{eqnarray}
\end{mathletters}
and since we can write
\begin{eqnarray}
f_i(z,w)=\rho_i(z)f_{z;i}(w)
\end{eqnarray}
where $f_{z}(w)$ is the vertical velocity probability distribution at height $z$, normalized so that $\int\!\!\, dw f_{z}(w) = 1$, we have
\begin{eqnarray}
\label{eq:hf2000}
\frac{\rho_{f_i}(z)}{\rho_i(0)} &=& \int_{-\infty}^{\infty}\! dw \, f_{z=0;i}\left(\sqrt{w^2 + 2\Phi(z)}\right).
\end{eqnarray}
By extracting the in-the-plane velocity distribution ${f}_{z=0}\left(w\right)$ from tracer populations of A and F-stars, and integrating to find $\rho_{\rm A,F}(z)$ for specific potentials $\Phi(z)$, Holmberg \& Flynn were able to test models for $\Phi$ by comparing the $\rho_{\rm A,F}(z)$ obtained in this fashion to the observed tracer densities. The model they considered was a modified version of the `Bahcall models' \citep{bahc84c,bahc84b,bahc84a}. The Bahcall model consists of splitting the visible matter into a series of isothermal components, each with a distinct vertical dispersion $\sigma_i$. By assuming dispersions and densities in the plane $\rho_i(0)$ for all components, the Poisson-Jeans equation (\ref{eq:poissonjeans1}) then gives a unique solution for $\Phi(z)$. The Bahcall model used by HF2000 is shown in Table 1 of the latter. An updated version of this model, featuring slight modifications, was used in \citet{hf2006}. 
Holmberg \& Flynn found that a mass model with little or no dark matter in the disk was in good agreement with the data, as did previously \citet{KG89b,KG89a} using a similar method. By adding or subtracting invisible mass to the various components in the model, HF then obtained a range on the acceptable mass models, which gave a range of acceptable densities as a function of height. Figure \ref{fig:hf2000} (top) compares the tracer densities with those predicted using the HF technique with an updated mass model containing no dark disk. We also show in Figure \ref{fig:hf2000} (bottom) the same result with a dark disk of surface density $\Sigma_D=10\mppc{2}$ and $h_D=10$ pc. Clearly it is of interest to know what is the maximum value of $\Sigma_D$ compatible with the data for a given $h_D$, within statistical errors.

\section{Sample}
\label{sec:method}
We now describe the numerical procedures used in the selection of our sample, as well as our methods for performing extinction corrections and for extracting the tracers' velocity distribution.

\subsection{Sample Selection}
\label{sec:sample}

We worked with the new reduction of the Hipparcos data \citep{hippnew}. As our sample, we used mostly the same stars as did HF2000, which contained A and F-stars. We performed the same cuts as HF2000. That is, for A-type stars, $-0.2<B-V<0.6$, and $0.0<M_V<1.0$, and the completeness limit $V\leq 7.9 + 1.1 \sin |b|$. By the completeness limit we have, using the definition of absolute magnitude,
\begin{eqnarray}
M_V+5 \log_{10} \left(\frac {d}{10 {\rm pc}}\right)&\leq 7.9 + 1.1 \sin |b|
\end{eqnarray}
and so
\begin{eqnarray}
\frac {d}{10 {\rm pc}}&\leq 10^{(7.9 + 1.1 \sin |b|-M_V)/5}
\end{eqnarray}
and since $|b|\ge 0$ and $M_V<1.0$, we have for our A-star sample, $d\lesssim 240\, {\rm pc}$ as our completeness limit. We therefore limited our sample to a vertical cylinder centered at the Sun with radius and half-height of $r_c=h_c=170$ pc. This ensures that the diagonals of the cylinder will be shorter than 240 pc. Note that HF2000 used a cylinder with radius 200 pc. The part of their A-star sample higher than 130 pc was therefore not complete. (By the same reasoning, their F-star sample was only complete to a height of 66 pc.) Using this method, our data set contained approximately 1500 A-stars. The precise numbers depend on the reddening corrections, as will be explained below.

For the F-star sample, the cuts in color were also $-0.2<B-V<0.6$, but here we used $1.0<M_V<2.5$. Here the maximum distance was found to be 120 pc. We therefore restricted our analysis to a cylinder of radius 50 pc and half-height of 109 pc. Additionally, to avoid any bias from the Coma Star Cluster \citep{coma}, we further cut off this sample from above at +40 pc. The final sample contained only about 500 stars, with the exact number depending on the extinction/reddening corrections and the value of the height of the Sun above the Galactic plane, $Z_\odot$.

\subsection{Extinction Corrections}
\label{sec:extinction}

HF2000 used the extinction model of \citet{hakkila}. We used the more recent 3D reddening map of \citet{schlafly}, which computes the reddening for most of the stars at a distance of 63 pc from the Sun or farther. For stars closer than this, we interpolated from 63 pc using the interpolation map of \citet{chen}, but with a sech${}^2$ profile for the reddening material instead of exponential. This interpolation model takes the dust scale height $h_{\rm dust}$ as an input. We assumed a sech${}^2$ scale height of 100 pc. For stars farther than 63 pc but not covered by the map, we extrapolated from the 2D dust map of \citet{finkbeiner}. 

\subsection{The Velocity Distributions}

For the vertical velocity of their stars, HF2000 used the approximation
\begin{eqnarray}
\label{eq:hfvel}
w = \frac{\kappa \mu_b}{\tilde{\pi} \cos b} + u \cos l\tan b + v \sin l \tan b.
\end{eqnarray}
where $\tilde{\pi}$ is the parallax and $\kappa=4.74047$ km$\cdot$s${}^{-1}\cdot$mas$\cdot($mas$\cdot$yr$)^{-1}$. This approximation is valid for stars with $\sin b \ll 1$. However, we make use of the exact equation
\begin{eqnarray}
\label{eq:krvel1}
w - w_0 = \frac{\kappa \mu_b}{\tilde{\pi}}\cos b + V_R \sin b
\end{eqnarray}
where $w_0=7.25$ km $\rm s^{-1}$ is the vertical velocity of the sun \citep{schonrich} relative to the LSR. Since radial velocities are not known, we simply  take an average:
\begin{eqnarray}
\label{eq:krvel2}
\langle w \rangle = w_0 + \frac{\kappa \mu_b}{\tilde\pi} \cos b + \langle V_R \rangle \sin b,
\end{eqnarray}
where we  take
\begin{eqnarray}
\langle V_R \rangle = -{\bf V}_{\rm sun}\cdot {\bf\hat{r}} = -u_0 \cos l \cos b - v_0 \sin l \cos b - w_0 \sin b
\end{eqnarray}
with $u_0 = 11.1$ km $\rm s^{-1}$ and $v_0=12.24$ km $\rm s^{-1}$, also given by \citet{schonrich}. This follows from the fact that from the LSR frame, we expect $\langle V^{\rm (LSR)}_R \rangle=0$ for any axisymmetric stellar population. Since we are also working with stars close to the plane ($\sin b\ll 1$), approximating $V_R$ by its average still gives a good approximation, and this approximation will still be better than Equation \ref{eq:hfvel}. This gives
\begin{eqnarray}
\label{eq:krvel3}
\langle w \rangle = \frac{\kappa \mu_b}{\tilde\pi} \cos b + w_0 \cos^2 b -u_0\cos l \cos b \sin b - v_0 \sin l \cos b \sin b.
\end{eqnarray}

We can do even better since radial velocities have been measured for many stars in the solar region since 2000. Therefore, rather than simply use Equation \ref{eq:krvel3} for all our stars, we supplemented the Hipparcos data with radial velocities from \citet{bbf}. We could thereby use measured radial velocities for 52\% of our stars within $|b|<12^\circ$. It was reported in \citet{binney97} that stars with measured radial velocities constitute a kinematically biased sample. However, \citet{korchagin} showed, for their red giant sample with measured $V_R$, the result only appears biased  when judging from their proper motions; once these are converted to cartesian $x$, $y$, and $z$-velocities, the kinematic bias is no longer observed. Moreover, we did not restrict our analysis to stars with measured radial velocities, since we completed the sample using the average $\langle V_R \rangle$. In light of these factors, along with the fact that we restrict our analysis to low $b$, we expect the radial velocities to introduce very little bias to our results.

\subsection{The Height of the Sun}
\label{sec:sun}
Because the Poisson-Jeans equation assumes thermal equilibrium, and because we have assumed Galactic azimuthal symmetry, we are considering only models with $z$-reflection symmetry across the Galactic plane; nonetheless, more general models are certainly possible. In these $z$-symmetric models, all disk-like components will be centered at the Galactic plane at $z=0$. In order to know the coordinates of the stars in our tracer populations, we need to know the height of the Sun relative to this Galactic midplane. Various measurements of this quantity have been made, but the precise result depends on the data set used. The value inferred from classical Cepheid variables is $Z_\odot = +\, 26 \pm 3$ pc \citep{Z0cepheid}. However, other measurements have found values as low as $Z_\odot = +6$ pc \citep{Z0}. In fact, for our tracer sample we find a best fit of $Z_\odot = +7\pm1$ pc. For values of $Z_\odot$ very different from this, the data in our sample were no longer consistent with any model for the Galactic potential. HF2000 also seem to have used a value of $Z_\odot$ very close to $Z_\odot\simeq0$. They may have been using the value $Z_\odot = +7$ pc, which they measured several years earlier in \citet*{HF97}. Interpreting this value in the context of DDDM implies that the Sun is currently within the dark disk, which could have important consequences for comet impacts \citep{matese,dino,rampino,shaviv}.

On the other hand, in Section \ref{sec:noneqintro}, we introduce the non-equilibrium version of the HF method. In this way of constraining the distribution, we can choose the initial position of the Sun, $Z_\odot$, in accordance with the measured value $Z_\odot = 26\pm 3$ or any other value since the population is assumed to be oscillating. The value affects the bound, because a larger initial value for $z$ for the stars will mean that they have a higher kinetic energy when they cross the midplane. This will cause them to spend a smaller fraction of their period interacting with the disk, predicting a weaker effect from a dark disk, and thus implying a looser bound. Conversely, a lower value for $Z_\odot$ will lead to a tighter constraint on the surface density $\Sigma_D$ of the dark matter disk. We will compute the bound assuming $Z_\odot=7\,{\rm pc}$ and $Z_\odot=26\, {\rm pc}$ separately. The currently favored value, $Z_\odot = 26\pm 3\,{\rm pc}$, being the highest measured value, should yield the most persistent bound. Assuming too small a height would yield a bound that could disappear if the Sun height is found to be bigger.

\section{Galactic Disk Components}

As previously stated, one of the main differences between our analysis and that of HF2000 is that we update the mass model for the Galactic disk in light of more recent observations. One important feature of this update is the use of midplane densities for the interstellar gas as inputs to the Poisson-Jeans equation. These were not available at the time of the HF2000 study. The gas surface densities can also be used to place further constraints on the dark disk by demanding self-consistency of the model. This is explained in a separate paper \citep{paper2} where we show how the combination of midplane and surface densities, which depend on the density profile in the vertical direction, can also set a bound on a dark matter disk. For now, we will discuss Holmberg \& Flynn's model and how we modify it. 

The Galactic disk model we update is \citeauthor{hf2006}'s slightly updated model from their 2006 study, which includes fainter stars than HF2000 and allows for the presence of a thick stellar disk. It is shown in Table \ref{tab:bahc2006}. In the far right column, we show the updated values, which we discuss in Sections \ref{sec:massmodel} and \ref{sec:stars}.

\begin{table}
\centering{}\caption{{ The Bahcall model used by \citet{hf2006}. The $\Sigma_i$ were calculated by \citeauthor{hf2006} from the solution to the Poisson-Jeans equation, except in the case of the interstellar gas, where they were held fixed by HF and the midplane densities were chosen to give the correct $\Sigma_i$. Note that revised values of $\rho_i(0)_{\rm new}$ give revised values of $\Sigma_i$ and that these are dependent on $\Sigma_D$.}}
\label{tab:bahc2006}
\begin{tabular}{rlccc|c}
\\
\tableline
$i$ & Description & $\rho_i(0)$ & $\sigma_i$    & $\Sigma_i$		& $\rho_i(0)_{\rm new}$\\
    &             &($\!\mppc{3})$ & (km s${}^{-2}$) & ($\!\mppc{2}$)	& ($\!\mppc{3}$)\\
\tableline
1 & H${}_2$ & 0.021  & 4.0  & 3.0 					&0.014*\\
2 & H${}_{\rm I}$(1) & 0.016  &7.0   & 4.1				&0.015*\\
3 & H${}_{\rm I}$(2) & 0.012  &9.0   & 4.1				&0.005*\\
4 & warm gas         & 0.0009 &40.0  & 2.0				&0.0011*\\
5 & giants	     & 0.0006 &20.0  & 0.4					&$0.0006^*$\\
6 & $M_V<2.5$	     &0.0031 &7.5   & 0.9				&0.0018\\
7 & $2.5<M_V<3.0$    &0.0015 &10.5  & 0.6				&$0^*$\\
8 & $3.0<M_V<4.0$    &0.0020 &14.0  & 1.1				&$0.0018^*$\\
9 & $4.0<M_V<5.0$    &0.0022 &18.0  & 1.7				&0.0029\\	
10& $5.0<M_V<8.0$    &0.007  &18.5  & 5.7				&0.0072\\
11& $M_V>8.0$        &0.0135 &18.5  & 10.9				&0.0216\\
12& white dwarfs&0.006  &20.0  & 5.4					&0.0056\\
13& brown dwarfs&0.002  &20.0  & 1.8					&0.0015\\
14& thick disk&0.0035  &37.0  & 7.0					&0.0035\\
15& stellar halo&0.0001  &100.0 & 0.6					&0.0001\\
\hline
\end{tabular}
 \\
 {\footnotesize * marks components whose dispersions $\sigma_i$ have been revised.}
\end{table}

\subsection{Insterstellar Gas}
\label{sec:massmodel}

Until now, one of the largest uncertainties in the visible mass model has been that of the interstellar medium. This is also the uncertainty most relevant to a dark disk, since it is the component most similar in scale height. The density and dispersion parameters used by HF2000 (and \citet{hf2006}, also quoted in \citealp{bt}) for the gas layer were taken from \citet{ismss} for the H${}_2$ layer and from \citet{ismkh} in the case of HI. Following Kulkarni \& Heiles, HF separate HI into the Cold Neutral Medium (CNM) and Warm Neutral Medium (WNM), due to their different scale heights. Based on the results of these studies, HF obtained a total gas surface density of about $13 \mppc{2}$, with an estimated uncertainty of  about 50\%.  In \citet{paper2}, we will show that these parameters have changed in recent years. This compilation, which is discussed in detail in \cite{paper2}, is cumbersome and tangential to our discussion here. We therefore simply  list the old and new values for the various disk components in Table \ref{tab:bahc2006} and leave the discussion of measurements to this separate paper, in which we also discuss how these measurements can be used independently to provide additional constraints on a dark disk model. We present the results with both the old and  the new values. We find that the corrections to the various parameters tend to compensate each other, so that, somewhat surprisingly, the new values do not change the results significantly.

The table shows the gas parameters including a factor of 1.4 \citep{ferriere} to account for the presence of helium and other elements. We also include the updated  dispersions for the interstellar gas components, based on a number of more recent studies, to 3.7 km $\rm s^{-1}$, 6.7 km $\rm s^{-1}$, and 13.1 km $\rm s^{-1}$, and $22 \;{\rm km\,s^{-1}}$ for ${\rm H_2}$, HI(1), HI(2), and HII respectively as explained in \citet{paper2}. We also include a contribution for thermal pressure, magnetic pressure, and cosmic ray pressure for HI(2) and HII as explained in \citeauthor{paper2}.

\subsection{Other Components}
\label{sec:stars}

 For the mass of the stellar components, we use the values reported by \citet{mckee}. These are shown in Table \ref{tab:bahc2006} (right column). Some of these are very different from those of \citet{hf2006}, such as the M-dwarf density (row 11). Differences in the other stellar components and in the gas components as well tend to compensate these changes so  that the total mass of the galactic disk is roughly equal to the HF value. The value `0' in row 7 reflects the fact that \citeauthor{mckee} grouped all the stars with $M_V<3$ into one category with a scale height given roughly by that of row 6. In addition to the midplane densities, we adjusted the dispersions of both the giants and the $3<M_V<4$ stars to 15.5 km $\rm s^{-1}$ and 12.0 km $\rm s^{-1}$ respectively to agree with the scale heights of \citet{mckee}. For the dark halo, we follow \citet[Equation 28]{bovytr}, who approximate the dark halo as a disk-like component with vertical dispersion $\sigma\simeq 130\,{\rm  km\, s^{-1}}$. Its midplane density was chosen so that $\rho_{\rm halo}(z=2.5 \,{\rm kpc})=0.008 \mppc{3}$ and depends on the specific values of $\Sigma_D$, $h_D$. We used this particular measurement because it relied on data high above the Galactic midplane and would be least biased by the existence or non-existence of a thin dark disk.

\subsection{Poisson-Jeans Solver}
We implement our Poisson-Jeans solver with the densities and dispersions as inputs, as did HF2000. Our solver is implemented by using Equation \ref{eq:poissonjeans2} in the following way. An initial distribution is assumed for each component $\rho_i^{(0)}(z)$. The potential $\Phi(z)$ due to these components is then calculated via
\begin{equation}
\Phi^{(0)}(z)=4\pi G\sum_i \int_0^{z}\!dz^{\prime}\,\int_0^{z^{\prime}}\!dz^{\prime\prime}\, \rho_i^{(0)}(z^{\prime\prime})
\end{equation}
which then gives the next iteration of $\rho_i(z)$:
\[\rho_i^{(N+1)}=\rho_i(0)\exp\left(-\Phi^{(N)}(z)/\sigma_i^2\right)\]
and this process is repeated until the solution converges. Remarkably, in only 5 iterations, the solution for a single component disk converges to the \citet{spitzer} solution (Equation \ref{eq:spitzersol}) to better than one part in $10^7$! This is affected only very slightly by the initial distributions $\rho_i^{(0)}$ assumed.

\section{Our Analysis}

\subsection{Traditional Method - Statistics}
\label{sec:stats}
In order to compare the stellar kinematics to a given dark disk model, we define  a $\chi^2$-type statistic $X$ that measures the distance between the predicted and observed densities:
\begin{eqnarray}
\label{eq:X}
X[\Phi]\equiv \int_{z_{\rm min}}^{z_{\rm max}}\!\!\!dz\; \frac{\Big|\rho_{\rm obs}(z)-\rho_{f,{\rm obs}}[\Phi(z)]\Big|^2}{\Delta^2_{\rho}(z)}
\end{eqnarray}
where
\begin{equation}
\rho_{f,{\rm obs}}[\Phi(z)] \equiv  \rho_{f}(z) = {\rho(0)}\int_{-\infty}^{\infty}\! dw \, f_{z=0}\left(\sqrt{w^2 + 2\Phi(z)}\right)
\end{equation}•
as defined in \ref{eq:hf2000}, $\Delta_\rho(z)$ represents the uncertainty in $\rho(z)$ at the position $z$, and where where $z_{\rm min},\;z_{\rm max}$ are given by the completeness limits, $Z_\odot\pm170 \;{\rm pc}$ for A stars and $Z_\odot-92\;{\rm pc}$, $Z_\odot+40\;{\rm pc}$ for F stars.

As explained in Appendix \ref{app:stats}, we can then bootstrap from the data to obtain the probability distribution $P(X)$ and obtain total probabilities on the A and F-star data $p(X_A,X_F|\Phi)$ given a dark matter disk model contained in the gravitational potential $\Phi$.

\subsection{A Cross-Check: Measuring $\Phi_\obs(z)$ Directly}
\label{sec:our}

It is worth pointing out that besides the traditional HF analysis which compares measured and model densities, it is also possible to use the HF equation to measure the Galactic potential $\Phi_\obs(z)$ directly from the data. This has the advantage that it treats the errors in density and velocity on more equal footing. This can be done by interpeting Equation \ref{eq:hf2000} as an equation for $\rho_{\rm A,F}(\Phi)$ (as in \citealt{KG89a}, Equation 20):
\begin{eqnarray}
\label{eq:rhoPhi}
\frac{\rho_{{f}}(\Phi)}{\rho(0)}=\int_{-\infty}^{\infty}\! dw_0 \, {f}_{z=0}\left(\sqrt{w_0^2 + 2\Phi}\right)
\end{eqnarray}
and then inverting this to obtain $\Phi_{{f}}(\rho)$. The `observed' gravitational potential is then given by combining this with the observed density ${\rho}_{\rm A,F}(z)$:
\begin{eqnarray}
\label{eq:our}
{\Phi}_\obs(z)\equiv \Phi_{{f}}\big({\rho_\obs}(z)\big)
\end{eqnarray}
Note that Equation \ref{eq:rhoPhi} always gives $\rho(\Phi)/\rho(0)\leq1$ for real $w$. For values of the observed density  ${\rho}_\obs(z)>\rho(0)$, we analytically continued $f_{z=0}(w)$ to imaginary $w$ (negative $\Phi$) by fitting $f_{z=0}(w)$ to a sum of three analytic Gaussians. Once we obtain ${\Phi}_\obs(z)$, we can then perform a cross-check on our analysis by comparing the measured ${\Phi}_\obs(z)$ with our model $\Phi(z)$ directly. The statistics can be accounted for similarly as in Appendix \ref{app:stats}, by sampling ${\Phi}_\obs(z)$ and computing their distribution.

\subsection{Non-Equilibrium Method}
\label{sec:noneqintro}
\begin{figure}[h]
\centering{}\caption{ (Left) The A-star velocity distribution possesses a peak value of $1.3 \pm 0.3$ km $\rm s^{-1}$. (Right) The A-star density distribution has a non-zero central value of $19\pm5$ pc relative to the Galactic plane, assuming a value for the solar position of $Z_0=26$ pc.}
\label{fig:noneqfeatures}
\plottwo{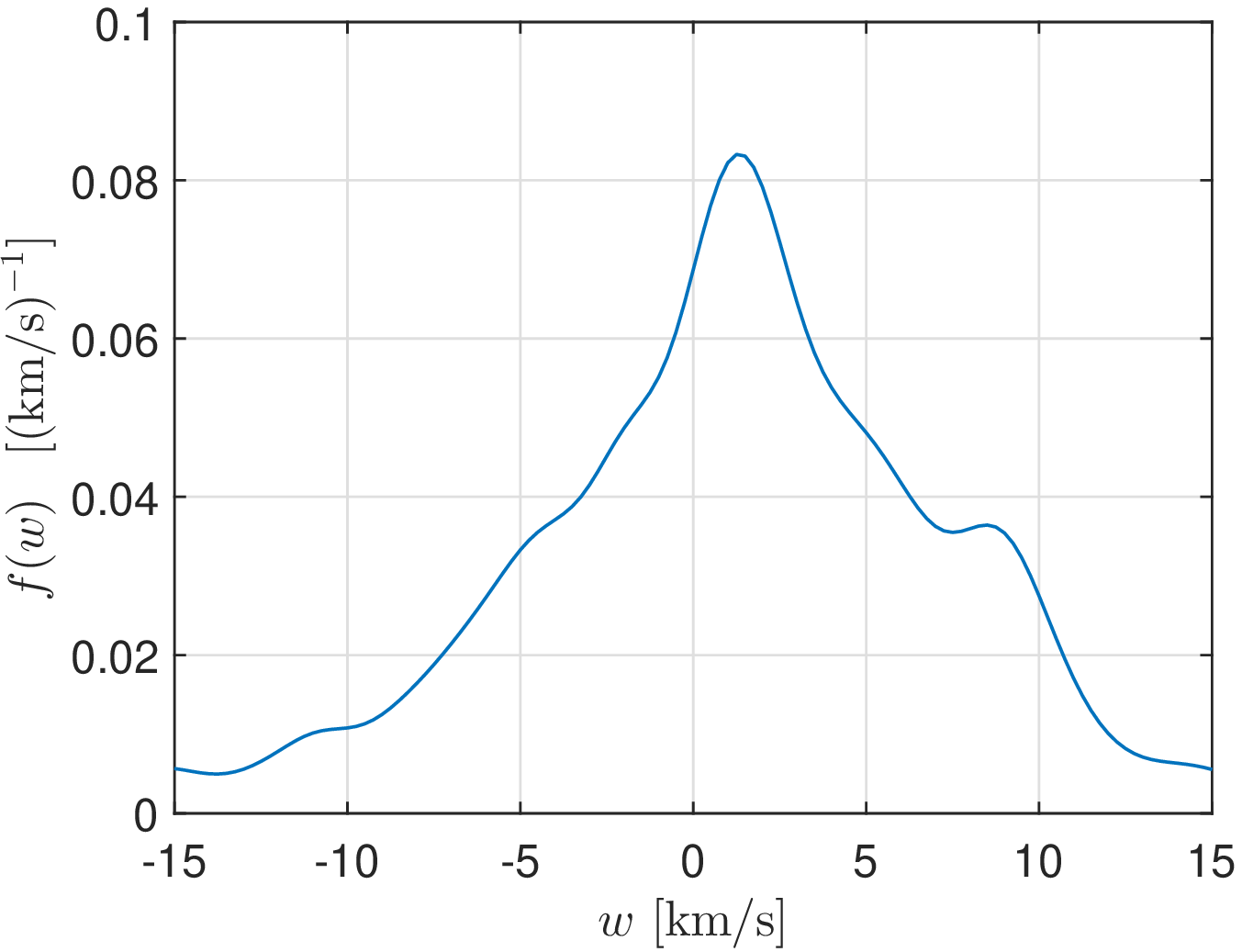}{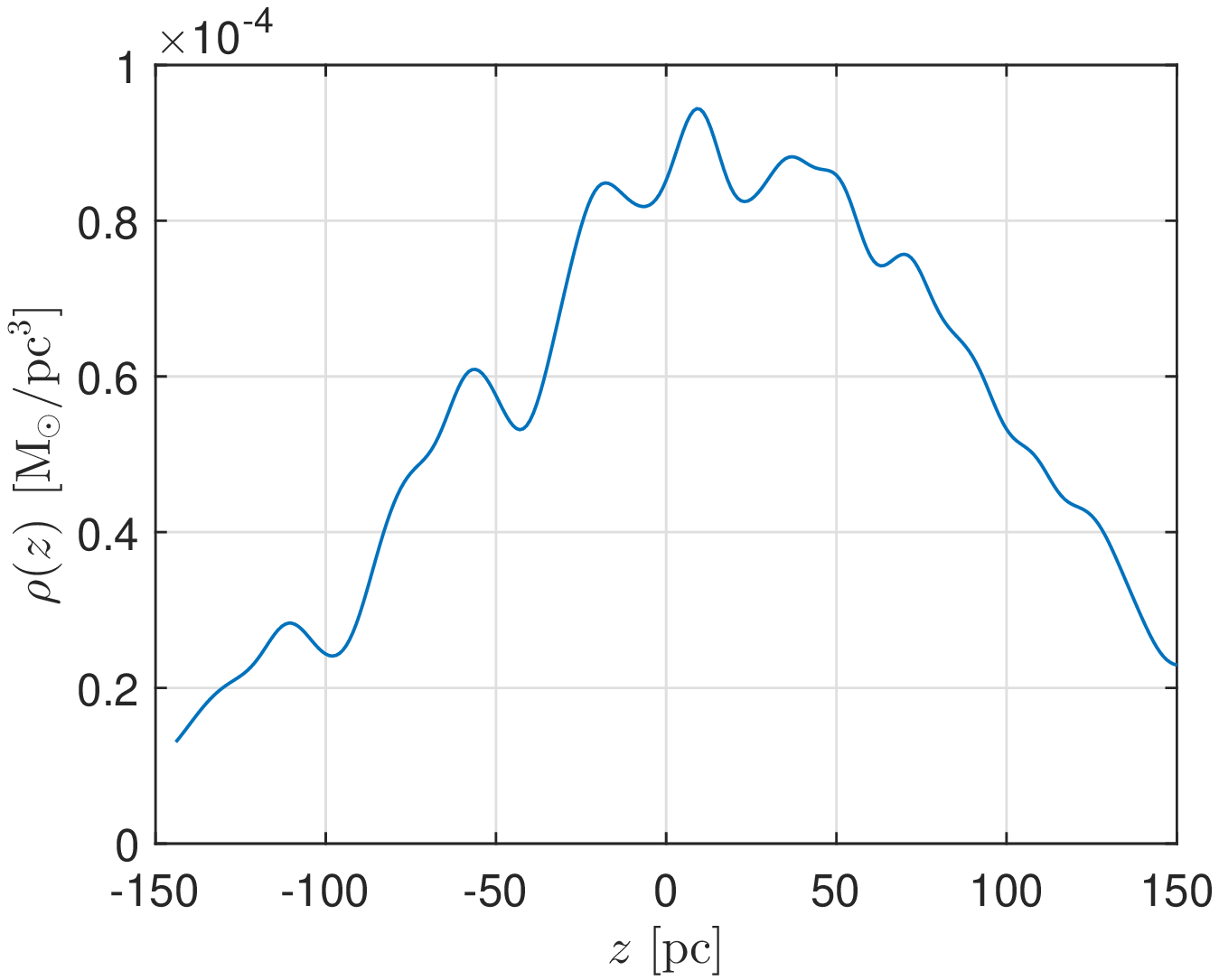}
\end{figure}

As was mentioned in Section \ref{sec:intro}, our tracer populations show evidence for deviations from equilibrium behavior. This evidence is a non-zero mean velocity and a displacement from the assumed position of the Galactic plane. Figure \ref{fig:noneqfeatures} shows the velocity and density distributions of the A stars. They feature a net velocity of $1.3\pm0.3$ km $\rm s^{-1}$ and a net displacement from the Galactic plane of $19\pm5$ pc. A bulk velocity and vertical displacement from the plane can clearly be seen.

Because of these features, the assumption ${\partial f}/{\partial t}=0$ in Equation \ref{eq:boltzmann} will not be satisfied.  Consequently, Equations \ref{eq:boltz} and \ref{eq:ez} no longer hold and the HF relation (\ref{eq:hf2000}) that in principle constrains the potential will no longer be useful for constraining the Galactic model. On the other hand, we might expect that, if the tracer distribution is oscillating about the plane in the Galactic potential, that the long-time average of ${\partial f}/{\partial t}$ will vanish. We therefore expect that the HF relation will hold for long-time averages. However, the only way to compute the long-time average of the star dynamics is to assume some form for the potential $\Phi(z)$. We show in Appendix \ref{sec:noneq} that in this case, the HF relation (Equation \ref{eq:hf2000}) is trivially satisfied for the potential $\Phi(z)$:
\begin{eqnarray}
\label{eq:hfav}
\overline{\rho(0)^{-1}}_{[\Phi]}\overline{\rho(z)}_{[\Phi]}=\int\!dw\;\left.\overline{{f_{z=0}}}_{[\Phi]}(\sqrt{w^2+2\Phi(z)})\right.
\end{eqnarray}
where $\overline{\phantom{(}\,\cdot\,\phantom{)}}_{[\Phi]}$ represents the time average under evolution in the potential $\Phi(z)$. Since Equation \ref{eq:hfav}, the correct statement of the HF relation, is trivially satisfied for any potential, it therefore cannot be used to constrain a dark matter model. Although we do indeed expect a dark disk to `pinch' the tracer distributions as explained in Section \ref{sec:toy}, this pinch cannot be measured with the HF method if the sample is oscillating. 

\begin{figure}[h]
\centering{}\caption{ (Left) Comparison of observed tracer distribution to distribution predicted with no dark disk. (Right) Comparison of observed tracer distribution to distribution predicted with dark disk of surface density $\Sigma_D=20\mppc{2}$, $h_D=10$ pc.}
\plottwo{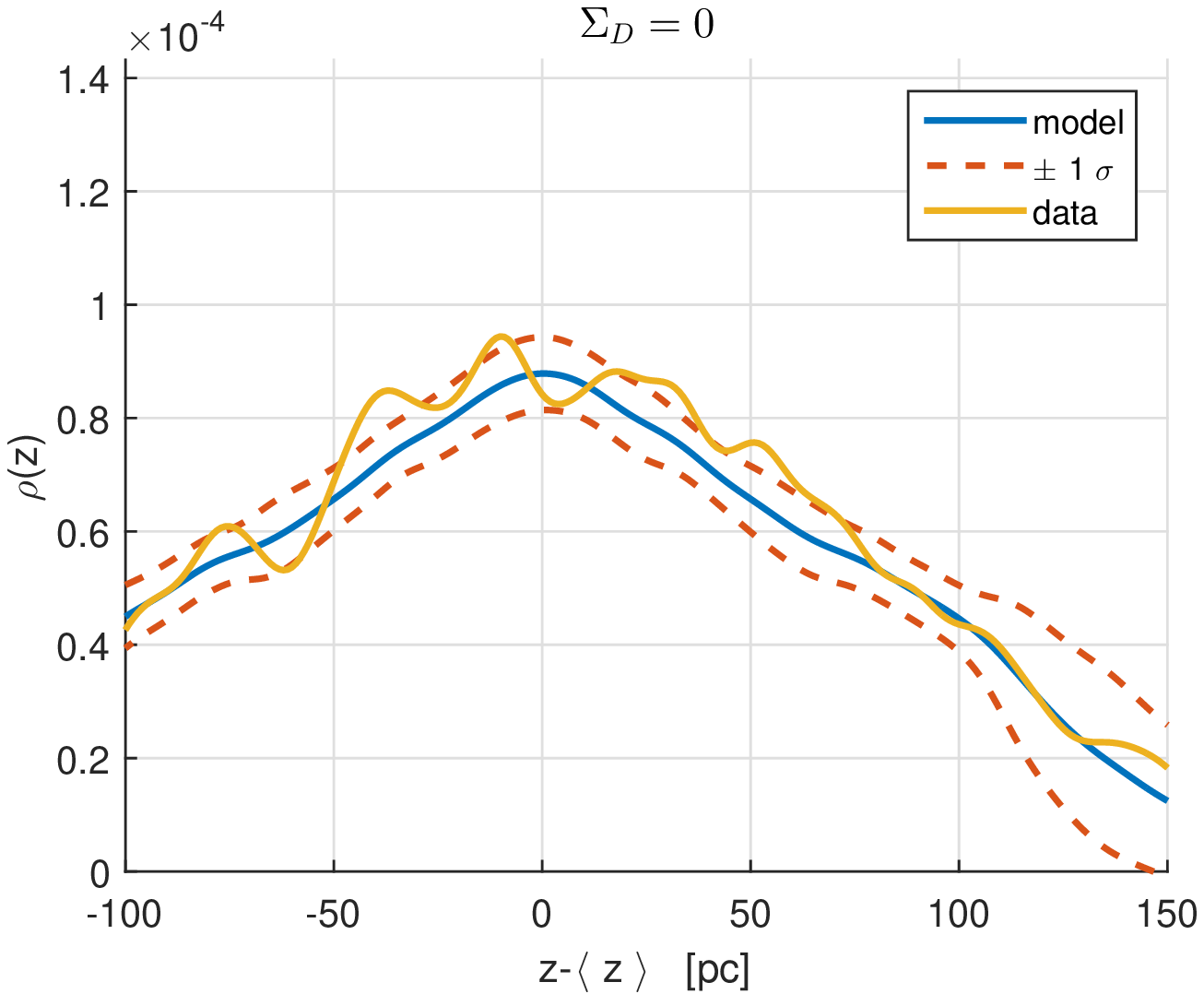}{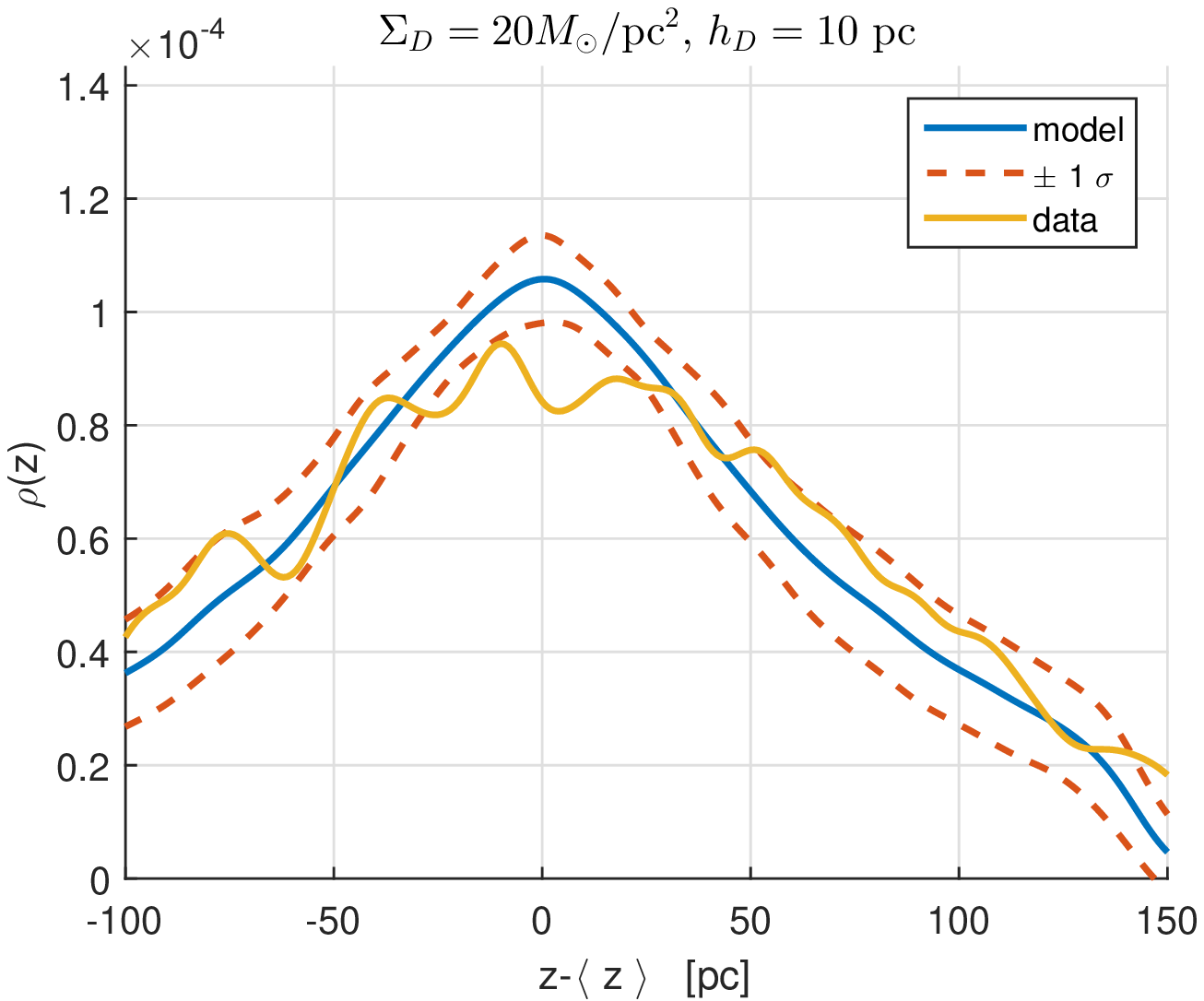}
\label{fig:tav}
\end{figure}

The solution we propose is to observe over time the shape of the distribution relative to the position of its center $z_0(t)$, i.e. $\rho(z-z_0(t),t)$, and to predict its time average, $\overline{\rho(z-z_0)}$. Figure \ref{fig:tav} shows this time average for a model with $\Sigma_D=0$ as well as with $\Sigma_D=20 \mppc{2}$. Comparing this to the Hipparcos `snapshot', which in our case is used as initial data, gives a method of constraining $\Sigma_D$ by assuming that the instantaneous distribution should not deviate strongly from its time average. We therefore define an appropriate $\chi^2$ parameter:
\begin{eqnarray}
\chi^2=\int\!dz\;\left|\frac{{\rho}_(z-z_0(0),0)-\overline{\rho(z-z_0)}}{\Delta_\rho(z)}\right|^2.
\end{eqnarray}
Where $\Delta^2_\rho(z)$ represents the expected variance in $\rho(z-z_0)$. For this, we use the variance $\Delta^2_\rho(z)$ computed in Section \ref{sec:stats}. This assumes that the distribution $\rho(z-z_0)$ is static in time, and that at any time, including the present $t=0$, we expect $\rho_{\rm obs}(z-z_0)$ to be equal to its average up to sampling error $\Delta_\rho(z)$. Although a more careful analysis would include the aforementioned sampling variance $\Delta^2_\rho$ as well as the variance due to the time dependence of $\rho(z-z_0)$, and will therefore be larger than $\Delta^2_\rho$, for the purposes of this analysis we neglected this contribution and considered only the sampling error $\Delta^2_\rho$. More careful determinations may be important in the future.

An alternate proposal for how the present distribution $\rho(z-z_0)$ fits the model is to measure the stability of the distribution over time. Were we to observe that the initial (possibly oscillating) distribution $\rho(z-z_0)$, decayed to a different (possibly oscillating) distribution, we would conclude that there was a dynamical mismatch between the distribution and the potential. On the other hand, were we to observe that the initial distribution retained its behavior over time, we would conclude that the distribution and potential were appropriately matched. Such an analysis would also have to account for spiral arm crossing and is beyond the scope of this manuscript. In any case, if evidence persists of nonstatic distributions, more careful analyses using nonstatic distributions will be required.

\section{Results and Discussion}
\label{sec:results}

\subsection*{Static Method}
We assigned probabilities to models with various $\Sigma_D$ and $h_D$ in the manner described in Section \ref{sec:stats}. The scale height $h_D$ was defined so that the value of the density of the dark disk at $z=h_D$ is $\rho(z=0)\;{\rm sech}^2(1/2)$.  
  
\begin{figure}[h]
\centering{}\caption{Standard HF analysis, without reddening corrections. $Top$: Relative probability density in DDDM parameter space as described in Appendix \ref{app:chi2}. $Bottom$: 95\% bounds on DDDM parameter space using conventional HF method for both A and F stars. The blue region is ruled out at $94\%$ to $95\%$ confidence. White regions are ruled out at $>95\%$. }
\plotone{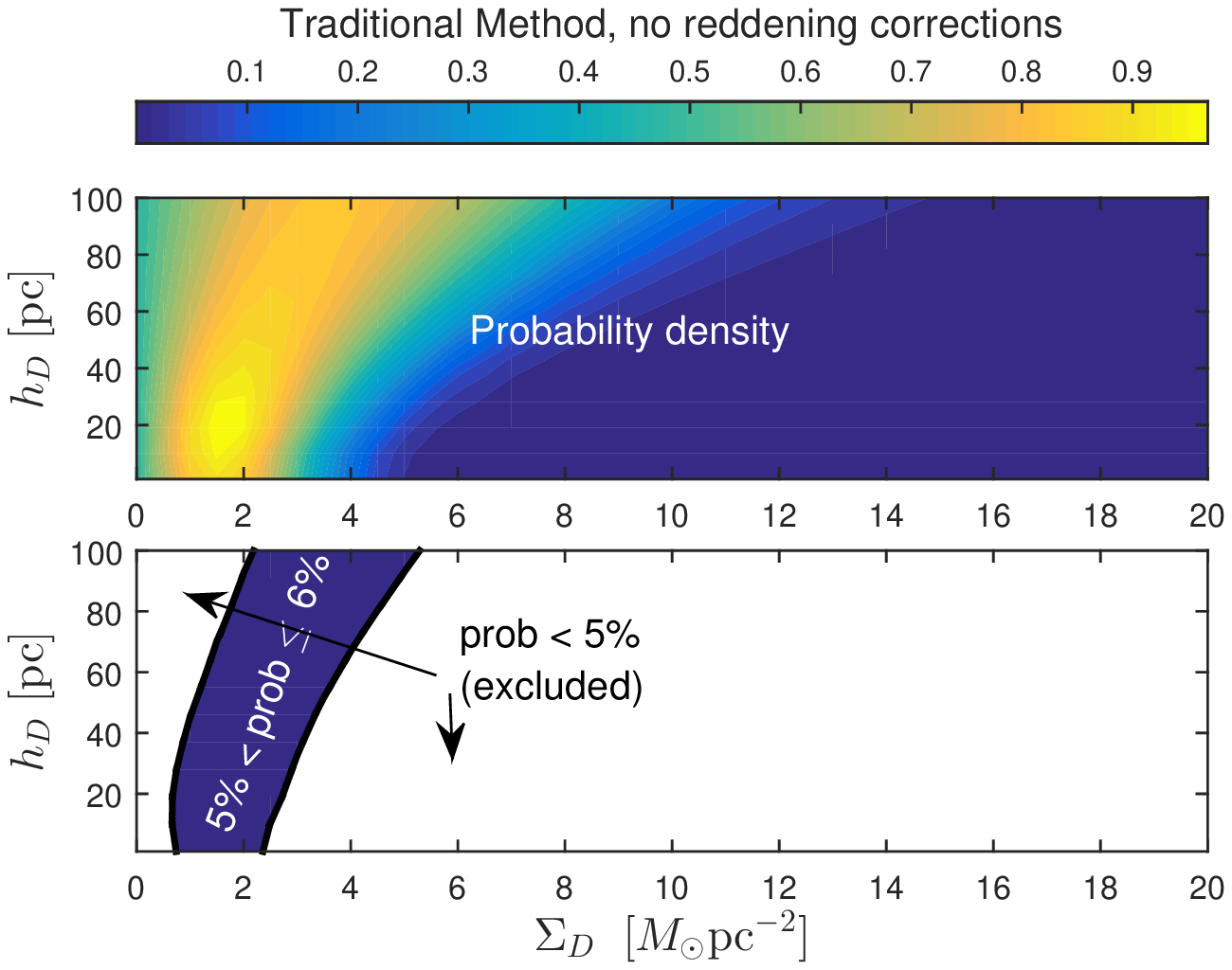}
\label{fig:nored}
\end{figure}

The probabilities and probability density derived without applying reddening corrections are shown in Figure \ref{fig:nored}. For low scale heights $h_D$, the probability has a 95\% upper bound at $\Sigma_D\simeq 2\mppc{2}$ and a 95\% lower bound of $1\mppc{2}$. For larger scale heights, the bound is slightly weaker, with the 95\% upper bound growing beyond $\Sigma_D\simeq 5 \mppc{2}$ at $h_D=80$ pc. However, without reddening corrections, the entire parameter space appears to be disfavored at 94\%. Moreover, without reddening corrections, the value $\Sigma_D=0$ appears to be excluded at greater than 95\% confidence. 

\begin{figure}[h]
\centering{}\caption{$Left$: Results of traditional HF analysis using mass model parameters of Flynn et al. (2006). $Right$: Results of traditional HF analysis using updated gas parameters of \citet{paper2}.}
\plottwo{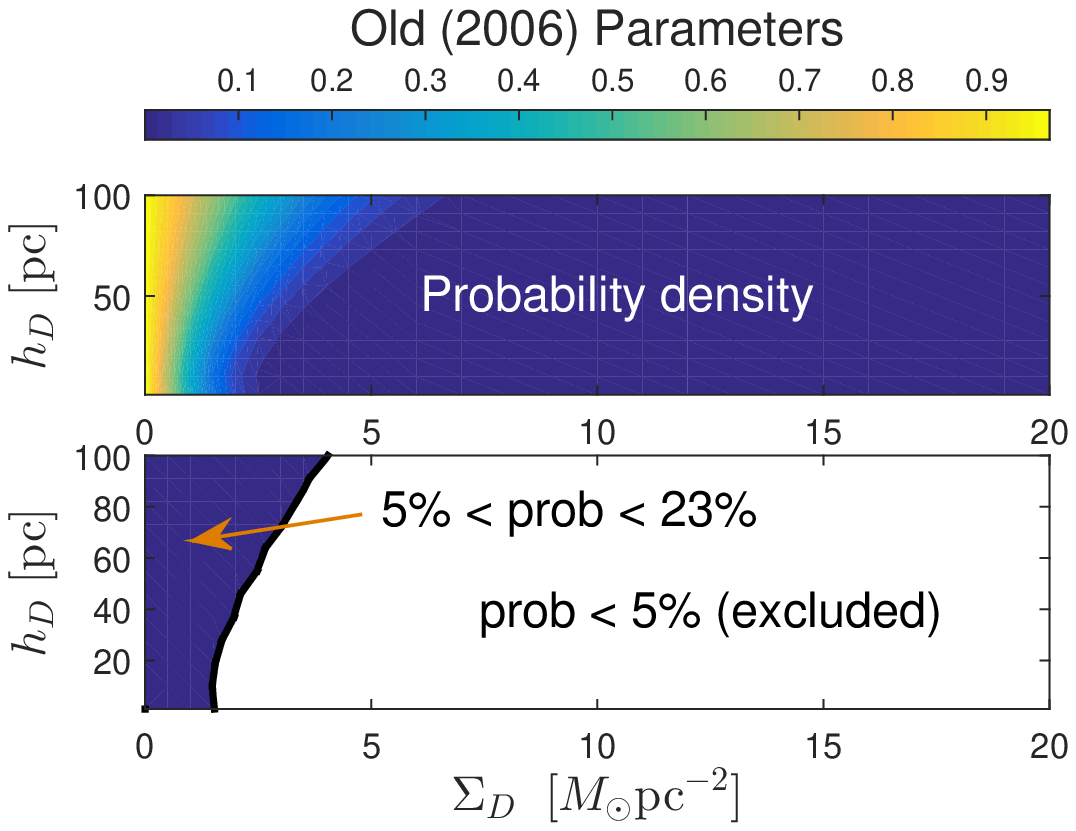}{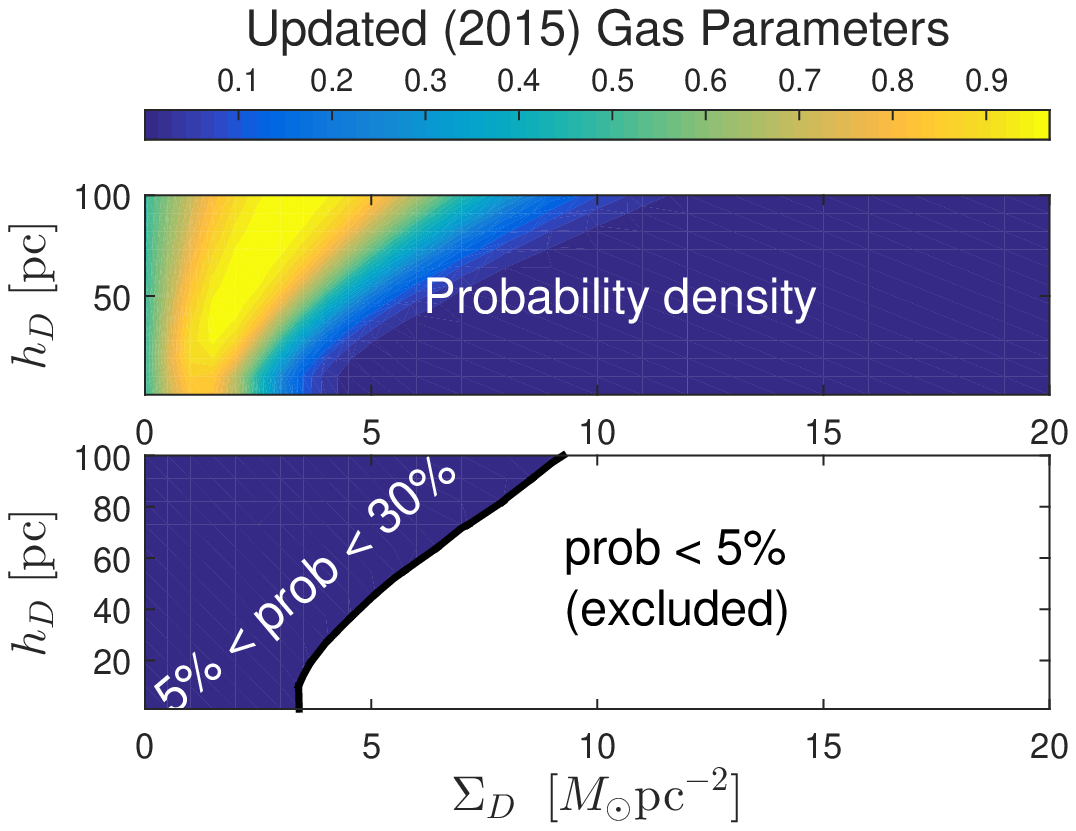}
\label{fig:results}
\end{figure}

\begin{figure}[h]
\centering{}\caption{$Left$: Results of traditional HF analysis using both updated gas parameters \citep{paper2} and updated stellar parameters \citep{mckee}. $Right$: 95\% bounds on DDDM parameter space using updated gas parameters one standard deviation below mean values.}
\label{fig:low}
\plottwo{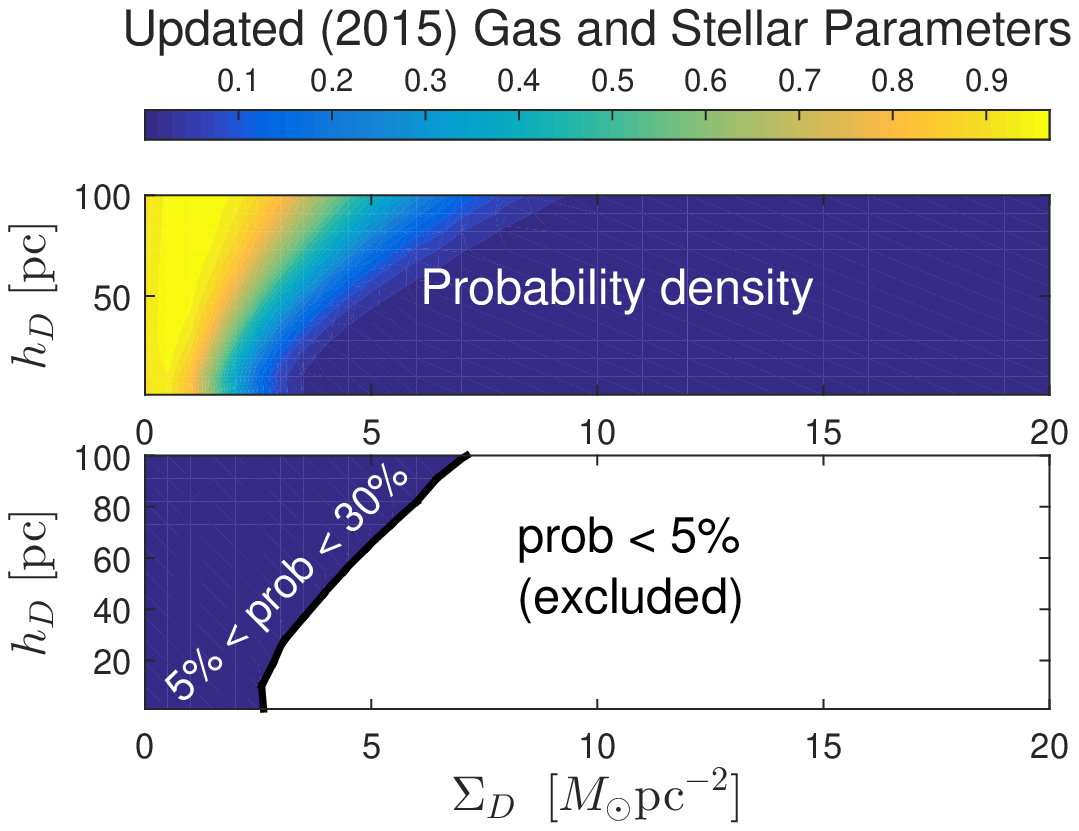}{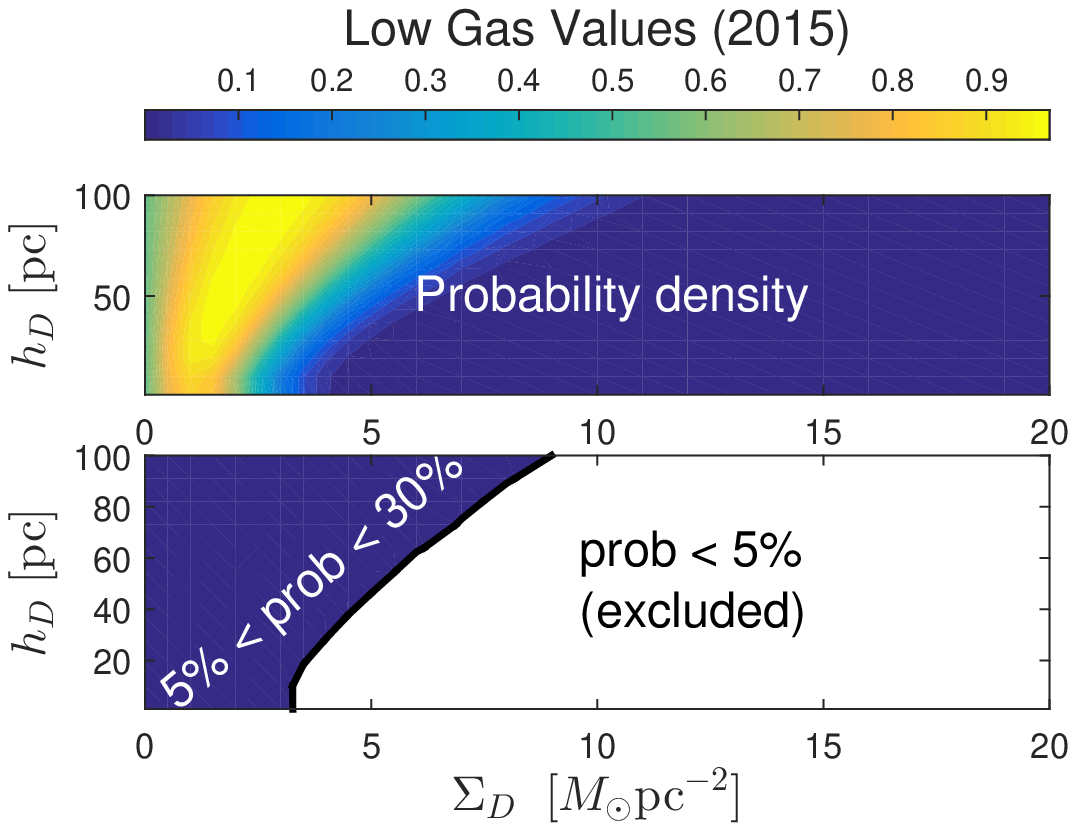}
\end{figure}

Figure \ref{fig:results}, on the other hand, shows, on the left, the results using the standard HF method used in Figure \ref{fig:nored} but with reddening corrections applied. This plot shows the results determined using the old values for the mass parameters from Flynn et al. (2006). We see that the upper bound is roughly the same with as without reddening corrections. Now, however, the entire parameter space is only disfavored at 77\%. On the right, we have the bounds using the new values for the gas parameters \citep{paper2} for comparison. The 95\% upper bound in this case (for low scale height) is around $4\mppc{2}$. Figure \ref{fig:low}, on the left, shows the values determined by using both the updated gas values and the updated values for the stellar components \citep{mckee}. Here, the bound is $3\mppc{2}$, which is remarkably similar to the bound using the 2006 values. This is because changes the changes in the various mass parameters tend to compensate each other on average, as would be expected statistically. 

In terms of midplane densities, the dark matter densities are less than 0.02$\mppc{3}$ for thick scale heights ($h_D\gtrsim 75\;{\rm pc}$) but are unbounded for low scale heights, varying as $h_D^{-1}$. Note that according to this analysis, the value $\Sigma_D=0$ is still ruled out at more than 70\% confidence, so one might still question the robustness of this method as applied to this data set and mass parameters.

In order to know whether the uncertainty in the interstellar gas mass parameters have an important effect on the bound, we compute the bounds using gas densities one standard deviation below their average values. The results are shown in Figure \ref{fig:low} on the right.  Here, the 95\% upper bound at low scale height ($h_D\sim10$ pc) is $3.5\mppc{2}$, which is slightly higher than the bound using the mean gas values. The value $\Sigma_D=0$ is still ruled out at more than 70\% confidence.

\begin{figure}[h] 
\centering{}\caption{68\% and 95\% bounds on DDDM parameter space using $\Phi(z)$ method.}
\label{fig:phi}
\plotone{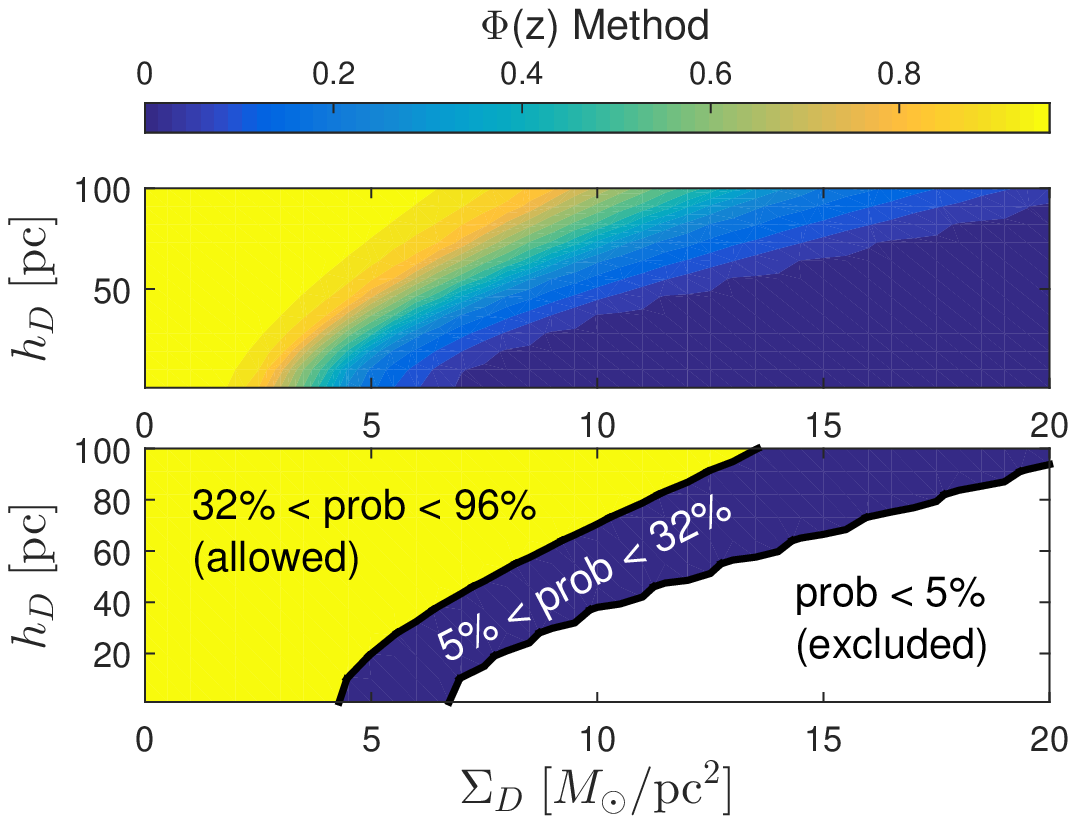}
\end{figure}

The plot in Figure \ref{fig:phi} shows the results of the cross-check mentioned in Section \ref{sec:our} where the HF equation was interpreted as an equation for $\Phi(\rho)$. The results of this method are consistent with the results from the standard analysis, although the bounds are weaker. This is because $\Phi(\rho)$ blows up as $\rho\to0^+$, causing large fluctuations in $\Phi(z)$ at high z and thus reducing the sensitivity of this method relative to the standard method.
 
\subsection*{Non-Equilibrium Constraint}
\begin{figure}[h]
\centering{}\caption{68\% and 95\% bounds on DDDM parameter space using the non-equilibrium version of the HF method for A stars only.}
\plotone{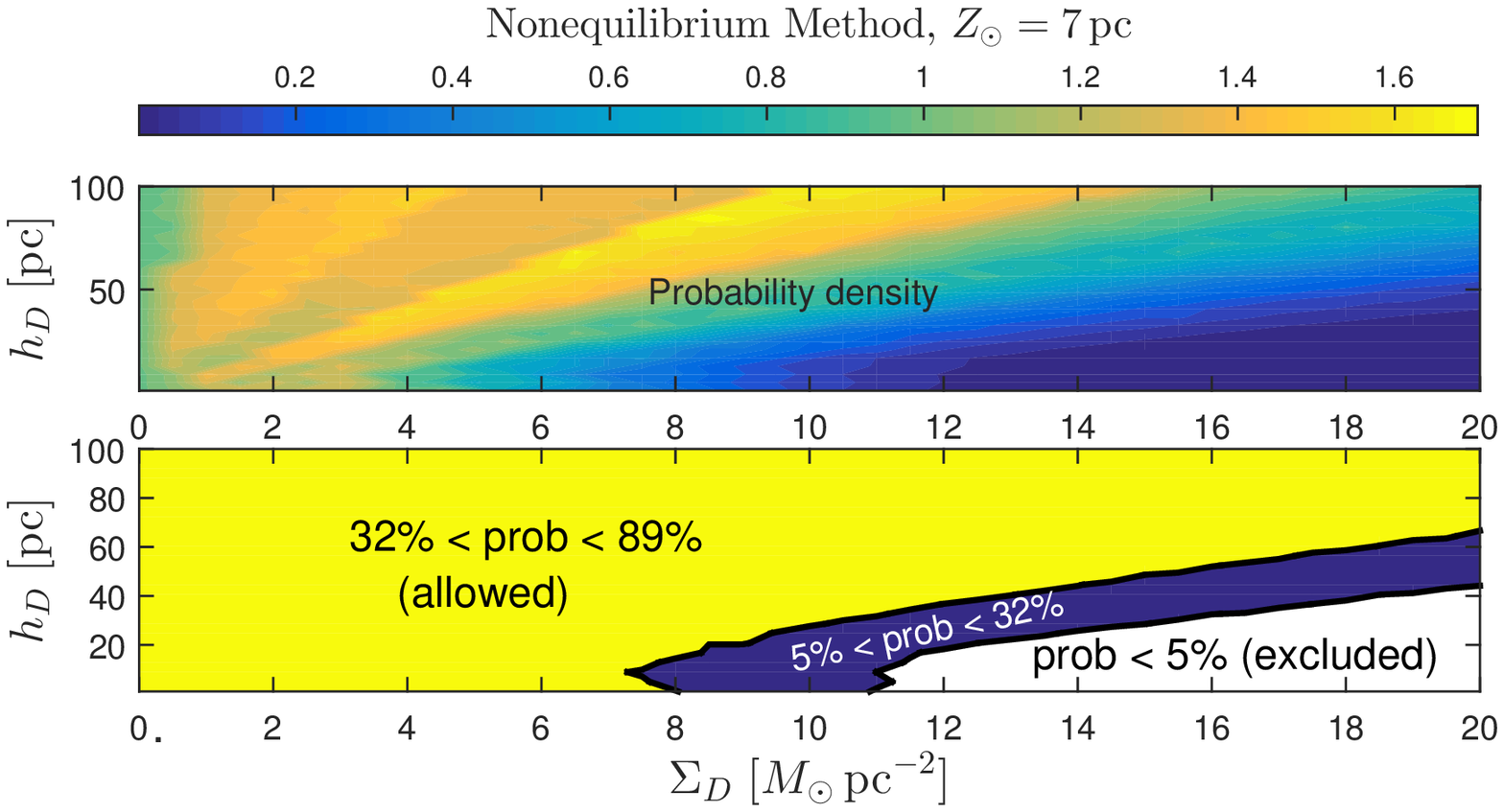}
\plotone{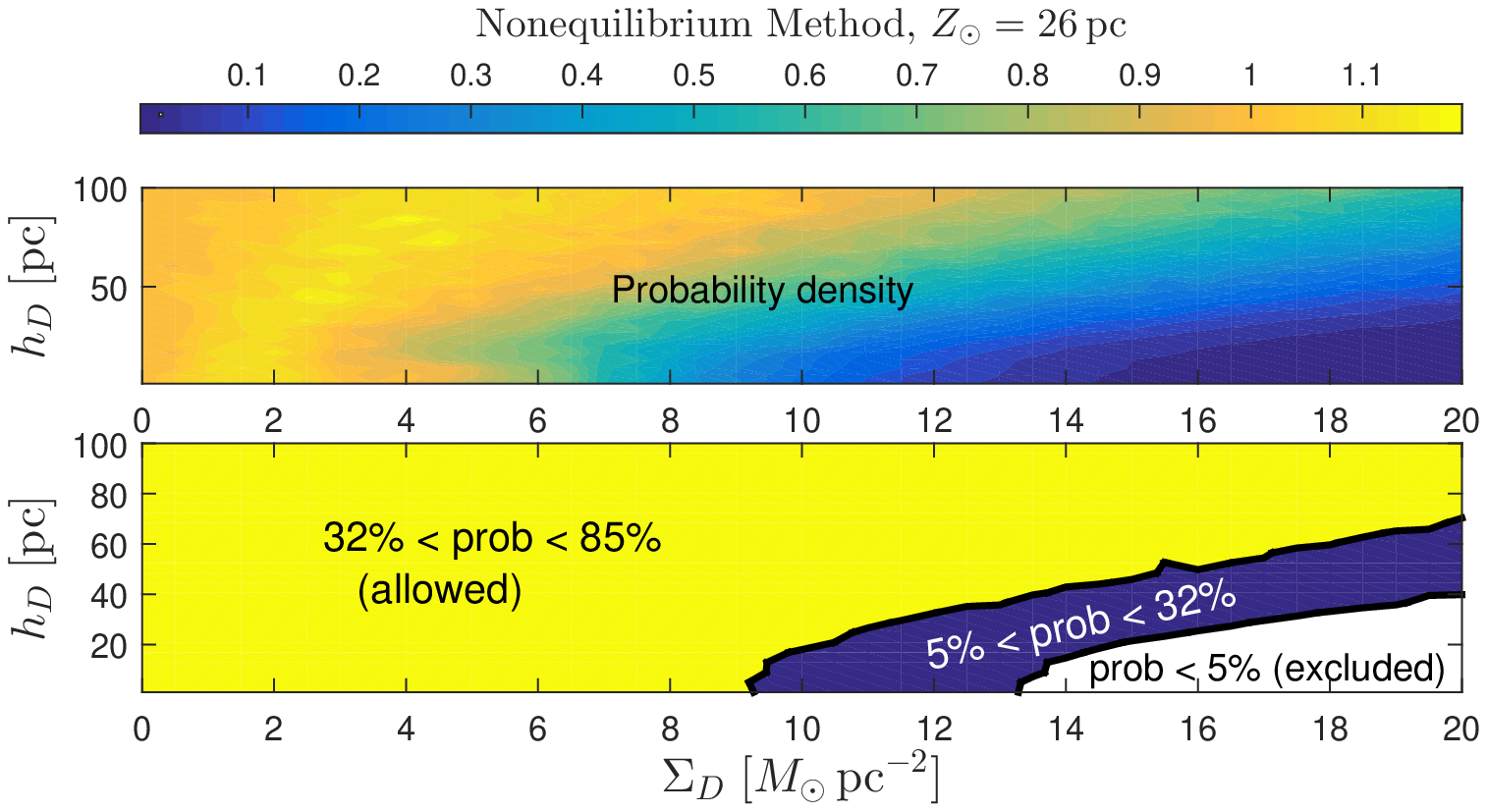}
\label{fig:noneq}
\end{figure}

Figure \ref{fig:noneq} shows the results obtained using the non-equilibrium HF method described in Section \ref{sec:noneqintro} and Appendix \ref{sec:noneq}, applied to the A star sample of Section \ref{sec:sample}. Here, we assume the distribution moves in the gravitational potential determined by the Galactic model and ask that the shape of the measured distribution not be far from that of the average distribution. The top plot shows the results computed assuming the low value of $Z_{\odot}=7$ pc ($cf.$ Section \ref{sec:sun}), i.e. assuming our distribution is centered at the Galactic plane. The bottom plot shows the results computed assuming the more accepted value of $Z_\odot=26$ pc. Note that the upper subplots in both figures show relative probability density, which, for $Z_\odot=26$ pc, very slightly seems to favor $\Sigma_D\simeq 2\mppc{2}$. In any case, the absolute probabilities clearly show that any values between $\Sigma_D=0$ and quite high density are allowed.

The  bound here is significantly weaker that of the static method. One reason is that the oscillations reduce the amount of time that the tracers spend in the dark disk, thus reducing their sensitivity to it. Applying the static method to such an oscillating population would not yield a bound on the parameter space since in this case, as explained in Section \ref{sec:noneqintro}, the static HF relation is trivially satisfied. Another reason the bound here is weaker simply depends on the amount of usable data in this method. This is reduced because the stars in the data are oscillating vertically, but in order to take a time average of the stars' distribution, we can include only values of $z$ that are covered by the data at $all$ times. In other words, the oscillation of the stars means that the $z$-cutoffs on the data are oscillating as well. The $z$-cutoffs on the time average are therefore the minimum values of these oscillating $z$-cutoffs. It is for this reason that the non-equilibirum analysis was perfored only on the A stars: the amount of F star data  available ($c.f.$ Section \ref{sec:sample}) after taking into account the oscillating $z$-cutoffs was not sufficient for analysis.

For low scale heights ($h_D$= 10 pc), in the non-equilibrium method, we find that the 95\% upper bound is between $\Sigma_D\simeq 10 \mppc{2}$ and $\Sigma_D\simeq 14 \mppc{2}$, and that the bound grows with $h_D$ as for the static case.

\subsection*{Galactic Disk Parameters}

\begin{figure}[h]
\centering{}\caption{Values of surface densities of the Galactic disk for a dark disk with scale height 10 pc. Black vertical dashed line corresponds to benchmark values $\Sigma_D=10 \mppc{2}$, $h_D=10\;{\rm pc}$. The NEQ dashed line corresponds to the current limits using the non-equilibrium method.}
\label{fig:surfacedens}
\includegraphics[scale=.8]{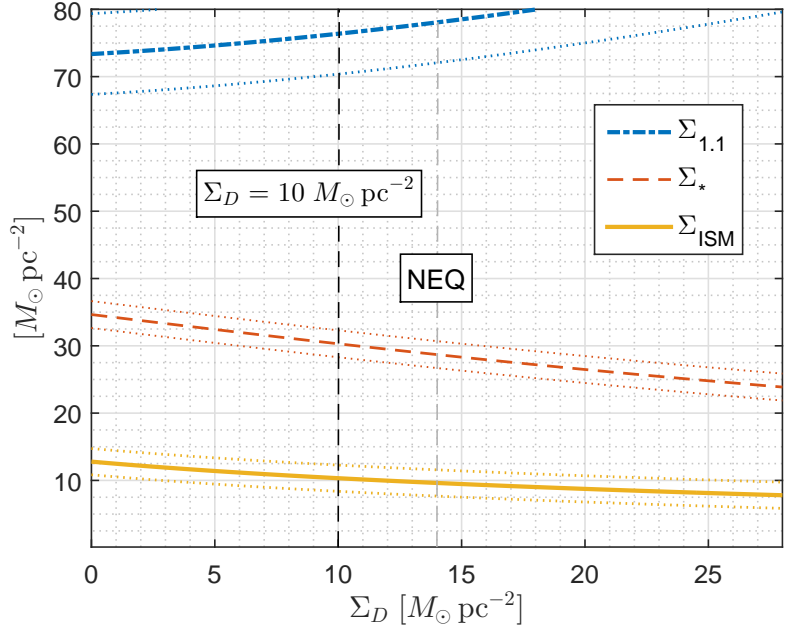}
\end{figure}

Another important result of our analysis is the value for the surface density of the Galactic disk, which we compare against existing measurements. Figure \ref{fig:surfacedens} shows the values of $\Sigma_{\rm ISM}$, the surface density of the interstellar medium; $\Sigma_*$, the surface density of the stellar disk (not including brown dwarfs and stellar remnants); and $\Sigma_{1.1}$ the total surface density to 1.1 kpc, including all visible and dark components (including e.g. brown dwarfs, dark halo, and dark disk). These were computed for a dark disk model with scale height 10 pc, and were obtained by integrating the self-consistent Poisson-Jeans solutions of each isothermal component. For the value of $\Sigma_D=10\mppc{2}$ at this scale height (black dashed line in Figure \ref{fig:surfacedens}), we find $\Sigma_{\rm ISM}\simeq 10\pm2 \mppc{2}$, the uncertainty being attributed to uncertainty in gas midplane densities, consistent with the value $\Sigma_{\rm ISM}=11.0\pm0.8\mppc{2}$ of \citet{paper2} and with the value $\Sigma_{\rm ISM}=12.8\pm1.5\mppc{2}$ of \citet{mckee}. At the $\Sigma_D$ upper bound obtained from the non-equilibrium method (gray dashed line), we have $\Sigma_{\rm ISM}=9.7\pm0.7\mppc{2}$. Values of $\Sigma_D$ higher than about $\Sigma_D=20\mppc{2}$ give lower values of $\Sigma_{\rm ISM}$ that are at odds with the literature. In \cite{paper2}, we perform a more detailed analysis of the self-consistency of the interstellar gas parameters and derive an independent bound on the DDDM parameter space.

 We also find $\Sigma_*=30\pm2\mppc{2}$ for the stellar surface density (not including brown dwarfs and stellar remnants) for $\Sigma_D=10\mppc{2}$, $h_D=10 {\rm pc}$. This agrees with the value $\Sigma_*=30\pm1\mppc{2}$ measured by \citet{bovydisk}. (On the other hand, the latter measurement was made assuming exponential profiles for the stellar disk components. Subleading corrections to this value obtained by assuming more realistic disk profiles near the plane give anywhere between 26 and 28$\mppc{2}$ depending on the specific model.) Note that our value was computed using the recent values of \citet{mckee}, and is somewhat higher than the values inferred using Holmberg \& Flynn's 2006 model. The latter's parameters yield $\Sigma_*=30\pm2\mppc{2}$ without the dark disk, and $\Sigma_*=26\pm2\mppc{2}$ when it is included. The uncertainties on our value of $\Sigma_*$ were estimated from those on the individual components given in \citeauthor{mckee} At the upper bound, we find $\Sigma_*=28.5\pm2.0\mppc{2}$. Note that $\Sigma_D=0$ gives $\Sigma_*=34\pm2\mppc{2}$, which is too large. The $\Sigma_*$ values therefore seem to favor $\Sigma_D\gtrsim 4 \mppc{2}$. 

For the total surface density to 1.1 kpc (still with $h_D=10$ pc), we find $\Sigma_{1.1}$ grows between $73.5\pm6.0\mppc{2}$ for $\Sigma_D=0$ and $\Sigma_D=76.5\pm6.0\mppc{2}$ for $\Sigma_D=10\mppc{2}$. These are slightly higher than the values in \citet{bovy14} who measured $68\pm4\mppc{2}$ but agree within their combined uncertainty. The major source of uncertainty in our measurement of this quantity is that of the dark halo density, which we set as described in Section \ref{sec:stars}. Here, too, we find that $\Sigma_D\gtrsim 20 \mppc{2}$ is at odds with the literatute. For thicker scale heights ($h_D\gtrsim50$ pc), we find that even for $\Sigma_D\gtrsim13\mppc{2}$, $\Sigma_{1.1}$ is too large.

\section{Conclusions}

 It is of interest to use existing and future kinematical data to ascertain the possible existence of a dark disk component in the Milky Way disk. Previous analyses argued that a dark disk is not necessary to match the data  but as has been often demonstrated, that is a far cry from ruling it out. Inspired by the Holmberg \& Flynn (2000) study, we have rederived the kinematic constraints on a dark disk by considering a dark disk in a self-consistent manner. Our analysis features updated kinematics and extinction corrections, an updated model for the interstellar gas, and careful statistics.
 
 We find that for a dark disk of sech${}^2$ scale height of 10 pc, the static method rules out surface densities greater than $3 \mppc{2}$ at 95\% confidence. In terms of midplane density, the favored value is $\rho_D = 0.0^{+ 0.1}\mppc{3}$, giving a total matter density in the plane of $\rho = 0.1^{+0.1}\mppc{3}$. These bounds increase with scale height. We note, however, that in the static method, even a model with $\Sigma_D=0$ is disfavored at around 70\%, pointing to the inadequacy of the method for the current data set. Using values of gas midplane densities a standard deviation lower than their average moves this bound closer to $4\mppc{2}$. We also find that the updated values of the mass model increase the bound relative to the old Flynn et al. (2006) values by about 50\% from $2\pm5\mppc{2}$ to $3\pm1\mppc{2}$.

These results were derived by arbitrarily removing the non-equilibrium features from our sample, which consisted of a bulk vertical velocity and a net vertical displacement from the Galactic midplane. However, if these features are instead taken into account using a non-equilibrium version of the HF analysis, the bound increases to $\Sigma_D\leq 14 \mppc{2}$, assuming $Z_\odot=26\,{\rm pc}$. This is partly because, in this method, it becomes more difficult to constrain a dark disk for the same amount of data. 

We also showed that the static HF method can be modified to directly measure the Galactic potential.

For thin dark disk models, the total surface density of the Galactic disk to 1.1 kpc, $\Sigma_{1.1}$, as well as the surface densities of visible matter, $\Sigma_*$ and $\Sigma_{\rm ISM}$, were found to be consistent with literature values both at the target values of $\Sigma_D$, $h_D$ and at their 95\% upper bound. According to our Poisson-Jeans solver, literature $\Sigma_*$ values appear to favor a dark disk model.
    
 Since the non-equilibrium analysis allows surface densities of low as zero and up to 14$\mppc{2}$, it follows that a dark disk may account for the comet periodicity that was extrapolated from the crater measurements. The results using the Gaia data promise to be more constraining, as more data close to the Galactic midplane will be available over a much larger area. However, it will be important to verify whether the sample used is in equilibrium. The correct method to use may indeed have to take account of the motion of the star populations, in which case non-equilibrium methods will be more appropriate.

\acknowledgements

We would like to thank Chris Flynn for all his comments and suggestions, and for reviewing our results. We would also like to thank Johann Holmberg for his comments. We would like to thank Jo Bovy for sharing his insights into the DDDM model and its constraints, and for pointing us towards the Holmberg \& Flynn technique for setting a bound on it, as well as for reviewing our results. We would also like to thank Chris McKee for sharing his most recent results with us and for reviewing our work. We thank our referee for useful suggestions and comments. We would like to thank Matt Reece and Doug Finkbeiner for discussions of the statistical analysis. Thanks also to Alexander Tielens, Katia Ferriere, and Matt Walker for help on the ISM parameters and to Chris Stubbs for his interest and comments. EDK was supported by NSF grants of LR and by Harvard FAS, Harvard Department of Physics, and Center for the Fundamental Laws of Nature. LR was supported by NSF grants PHY-0855591 and PHY-1216270. Calculations were performed using MATLAB 2015a.

\appendix
\section{Appendix - Statistics}
\label{app:stats}

In order to compare the stellar kinematics to a given dark disk model, we define  a $\chi^2$-type statistic $X$ that measures the distance between the predicted and observed densities:
\begin{eqnarray}
\label{eq:X}
X[\Phi]\equiv \int_{z_{\rm min}}^{z_{\rm max}}\!\!\!dz\; \frac{\Big|\rho_{\rm obs}(z)-\rho_{f,{\rm obs}}[\Phi(z)]\Big|^2}{\Delta^2_{\rho}(z)}
\end{eqnarray}
where
\begin{equation}
\rho_{f,{\rm obs}}[\Phi(z)] \equiv  \rho_{f}(z) = {\rho(0)}\int_{-\infty}^{\infty}\! dw \, f_{z=0}\left(\sqrt{w^2 + 2\Phi(z)}\right)
\end{equation}•
as defined in \ref{eq:hf2000}, $\Delta_\rho(z)$ represents the uncertainty in $\rho(z)$ at the position $z$, and where where $z_{\rm min},\;z_{\rm max}$ are given by the completeness limits, $Z_\odot\pm170 \;{\rm pc}$ for A stars and $Z_\odot-92\;{\rm pc}$, $Z_\odot+40\;{\rm pc}$ for F stars.  As explained in Section \ref{sec:sample}, for the F stars, we do not include any data higher than 40 pc above the Sun.

The observed densities $\rho_{\rm obs}(z)\equiv\rho_{\rm A,F}(z)$ were constructed using the kernel histogram technique. 
That is, we represented each star as a Gaussian with unit area in position and velocity space, and then obtained total distributions by summing these Gaussians. The uncertainties in position and velocity of the individual stars, $\Delta z$ and $\Delta w$, do not however fully acount for the error. The chief source of error on this sparse distribution is the  likelihood that stars wll actually fill in the purported distribution (i.e. Poisson error). 

We estimate the latter in the following way. If, in the case of $\rho(z)$, we assume the data are arranged in `bins' of half-width $\Delta z$, then the density $\rho(z)$ should be
\begin{equation}
\rho(z) = \frac{N(z) }{2A\Delta z}
\end{equation}
where $N(z)$ is the number of stars in each `bin' and $A$ is the cross-sectional area of the bin at height $z$. Both $\Delta z$ and $N(z)$ are unknown. However, we know that the fluctuations in $\rho(z)$ should be given, according to Poisson statistics, by
\begin{equation}
\Delta_\rho(z) = \frac{\sqrt{N(z)}}{2A\Delta z}
\end{equation}
Eliminating $N(z)$, we have
\begin{equation}
\label{eq:dzpoisson}
\Delta z = \frac{\rho}{2A\Delta^2_\rho}.
\end{equation}•
Since the determination of $\rho(z)$ and $\Delta_\rho(z)$ themselves depend on the value of $\Delta z$, Equation \ref{eq:dzpoisson} should be used recursively. Also, since the derivation of Equation \ref{eq:dzpoisson} was heuristic, we estimate this computation of $\Delta z$ to be correct only to within a factor of two or so. In this way, we obtain values of $\Delta z$ in the range 6-12 pc. To this we must add the $\Delta z$ arising from measurement error. This is derived by propagating the error in parallax available in the Hipparcos catalogue and grows with $z^2$. Summing these two contributions in quadrature gives values of $\Delta z$ between $\Delta z \simeq 7\, {\rm pc}$ near $z=0$ and $\Delta z \simeq 20\,{\rm pc}$ near the extremeties in the case of A stars, and $\Delta z \simeq 7-13\,{\rm pc}$ for the F stars. On the other hand, standard kernel density estimation techniques (which assume a constant kernel width and unimodal distribution) suggest widths closer to \citep{silverman}:
\begin{equation}
\Delta z \simeq \frac{{\rm stdev}\{z_i\}}{N^{1/5}}\simeq 18\, {\rm pc}
\end{equation}
and
\begin{equation}
\Delta w \simeq \frac{{\rm stdev}\{w_i\}}{N^{1/5}}\simeq 2.1\, {\rm pc},
\end{equation}
for A stars. We find similar values for the F stars. We find that the final result is roughly independent of this width for the range $\Delta z \simeq 6-30$ pc. The weakest bound is obtained for $\Delta z\simeq 18$ pc, before the width of the kernels becomes so large that it biases the distribution and begins to remove the signature of any potential structure near $z=0$.

In order to construct the relevant uncertainty $\Delta_\rho(z)$, we considered the density distributions $\rho_{\rm obs}(z)$ obtained by repeatedly sampling a fraction $q<1$ of the stars in the respective samples. We did this by either including or not including each star in the original data set with probability $q$. The variance between these distributions thus gives the spread $\Delta^{(q)}_{\rho}(z)$ at every point $z$. In terms of $\Delta^{(q=1/2)}_{\rho}(z)$ obtained from repeatedly sampling half the stars in the sample, we can obtain the uncertainty $\Delta_{\rho}(z)$ on the original sample as (Equation \ref{eq:factor2})
\begin{eqnarray}
\label{eq:sigma12}
\Delta_{\rho,{\rm obs}}(z)=2\,\Delta^{(q=1/2)}_{\rho}(z)
\end{eqnarray}
which can be derived using the binomial distribution with probability $q$ (see Appendix \ref{sec:q}). In numerical simulations, we find this factor of 2 is closer to 1.97. Although we do not know the source of this discrepancy, we note that its effect is small compared to the remaining errors in the analysis. The errors $\Delta_{\rho_{f,\obs}}$ were constructed in a similar way by sampling velocities from the data set, giving (Equation \ref{eq:factor1})
\begin{eqnarray}
\label{eq:sigma12}
\Delta_{\rho_f,{\rm obs}}(z)=\Delta^{(q=1/2)}_{\rho_f}(z).
\end{eqnarray}
The errors were then added in quadrature:
\begin{eqnarray}
\Delta^2_\rho(z)=\Delta^2_{\rho,{\rm obs}}(z)+\Delta^2_{\rho_{f,\obs}}(z).
\end{eqnarray}•
 Uncertainties in the Galactic potential, obtained from Sections \ref{sec:massmodel} in the case of interstellar gas and from \citeauthor{mckee} in the case of stellar components were were found to be negligible compared to the errors above. 
 
 In order to associate a value of $X[\Phi]$ with a probability, we constructed a probability distribution for the values of $X[\Phi]$ obtained from fluctuations of the density and velocity distributions. To do this, we note that, given the true potential $\Phi_{\rm true}$ of the Galaxy, the value
\begin{eqnarray}
X[\Phi_{\rm true}]=\int\! dz \frac{\left|\rho_{\rm obs} -\rho_{f,{\rm obs}}[\Phi_{\rm true}]\right|^2}{\Delta_{\rho}^2}
\end{eqnarray}•
is itself a fluctuation with respect to its equilibrium value 
\begin{eqnarray}
X_{\rm eq}[\Phi_{\rm true}]=\int\! dz \frac{|\rho_{\rm eq} - \rho_{f,{\rm eq}}[\Phi_{\rm true}]|^2}{\Delta_\rho^2}=0,
\end{eqnarray}• as $\rho_{\rm obs}$ and $f_{\rm obs}$ are assumed to be fluctuations of their equilibrium values $\rho_{\rm eq}$, $f_{\rm eq}$. We can similarly define $\Phi_{\rm obs}$ so that
\begin{eqnarray}
\rho_{f,{\rm obs}}[\Phi_{\rm obs}]=\rho_{\rm obs}
\end{eqnarray}•
and $X[\Phi_{\rm obs}]=0$. We then compute fluctuations in $X[\Phi_{\rm obs}]$ by sampling fluctuations in $\rho_{\rm obs}$ and $f_{\rm obs}$. That is, each time $k$ that we sample a set $\{z^\prime,w^\prime\}_k$ from the parent populations $\{z,w\}$, this gives an observed density $\rho_{{\rm obs},k}(z)$ and $f_{{\rm obs},k}(w)$. We therefore have
\begin{eqnarray}
\label{eq:deltaX0}
X_k[\Phi_{\rm obs}] = \int\!dz \frac{\left|\rho_{{\rm obs},k} -\rho_{f,{\rm obs},k}[\Phi_{\rm obs}]\right|^2}{\Delta_{\rho}^2}.
\end{eqnarray}•
Thus, by repeatedly sampling values of $X_k$, we obtain the distribution $P(X)$. We can then use the distribution $P(X)$ to associate probabilities with every point in parameter space by computing the cumulative $X$ distribution:
\begin{eqnarray}
C(X)=\int_0^X\!\! dX^\prime\; P(X^\prime).
\end{eqnarray}•
$C(X[\Phi])$ represents the probability that the fluctuations in $\rho_{\rm obs}$, $\rho_{f,\obs}[\Phi]$ with respect to the equilibrium distributions $\rho_{\rm eq}$, $f_{\rm eq}$ assuming that $\Phi_{\rm true}=\Phi$ could result in a value $X\leq X[\Phi]$. The probability $1-C(X[\Phi])$ therefore represents the the probability that $X\geq X[\Phi]$. In Bayesian terms, this is the probability $p(X|\Phi)$. Combining the independent results from the A and F stars, we have the probability
\begin{eqnarray}
p(X_A,X_F|\Phi)=(1-C(X_A[\Phi]))\times(1-C(X_F[\Phi]))
\end{eqnarray}•
We reject any model for which $p(X_A,X_F|\Phi)<0.05$. This will be our criterion for ``exclusion at 95\% confidence''. Note that this does not represent the probability of the model itself, but rather the probability of the results given the model.
  
As an aside, probability densities in model space $\Sigma_D$, $h_D$ can also be computed by fitting the $P(X)$ distribution to a $\chi^2$ distribution, as explained in Appendix \ref{app:chi2}. However, since this fit involves two slightly degenerate parameters, as explained therein, these probability densities are only approximate and are included in our results for illustration purposes only. We will see that, even so, they match qualitastively what is suggested by the absolute probabilities.

\section{Appendix - Constructing Errors}
\label{sec:q}

In Appendix \ref{app:stats}, we claimed that we could construct the errors on the total distributions $\Delta_\rho(z)$ or $\Delta_f(w)$ by repeatedly sampling a fraction $q$ of the stars. We now proceed to derive the relationship between these values and the values $\Delta^{(q)}_{\rho}(z)$, $\Delta^{(q)}_{f}(w)$ obtained by sampling.

We model the positions and velocities $\{z_i,w_i\}$ of the stars as being drawn from some `parent' probability distribution $f(z,w)$. If we sample a very small fraction $q\ll1$ of the stars, we expect the randomness in the resulting set $\{z^\prime_{i^\prime},w^\prime_{i^\prime}\}$ relative to the set ${z_i,w_i}$ to mimic the randomness of the set $\{z_i,w_i\}$ relative to the parent probability distribution $f(z,w)$, with the the errors reduced by a Poisson factor $\sqrt{q}$. We can therefore estimate the randomness in the parent set $\{z_i,w_i\}$ by repeatedly sampling a small fraction $q$ from this set and dividing the standard deviation $\Delta^{(q)}$ in the resulting set by $\sqrt{q}$:
\begin{eqnarray}
\label{eq:limq}
\Delta = \lim_{q\to0} \frac{\Delta^{(q)}}{\sqrt{q}}.
\end{eqnarray}
However, since the data set $\{z_i,w_i\}$ is of finite size, for $q\ll1$ we cannot generate a large enough child set $\{z^\prime_{i^\prime},w^\prime_{i^\prime}\}$ to be able to obtain reliable statistics. On the other hand, if we use a larger sampling fraction $q$, Equation \ref{eq:limq} will no longer hold because different samplings $i^\prime$ will repeatedly contain the same values. The trick will therefore be to relate $ \lim_{q\to0} {\Delta^{(q)}}/{\sqrt{q}}$ to a reference value $\Delta^{(q_0)}$ for some reference $q_0=\mathcal{O}(1)$. 

In order to derive the dependence of $\Delta^{(q)}$ on $q$, consider binning the data $\{z_i,w_i\}$ into bins in $z$ and $w$ space. For concreteness, let us consider only $z$ space for now. Let $N_k$ be the number of stars in bin $k$. Each star in bin $k$ has a probability $q$ of being sampled, and a probability $1-q$ of not being sampled. The distribution of outcomes is therefore binomial:
\begin{mathletters}
\begin{eqnarray}
1&=&(q+1-q)^{N_k}\\
&=&\sum_{r=0}^{N_k}\binom{N_k}{r}q^r(1-q)^{N_k-r},
\end{eqnarray}
\end{mathletters}
in the sense that the probability of sampling $r$ stars in the bin $k$ is given by 
\begin{mathletters}
\begin{eqnarray}
P_{k,r}&=&\binom{N_k}{r}q^r(1-q)^{N_k-r}\\
&=&(1-q)^{N_k}\binom{N_k}{r}\left(\frac{q}{1-q}\right)^r\\
&\equiv&(1-q)^{N_k}\binom{N_k}{r}\tilde{q}\,{}^r
\end{eqnarray}•
\end{mathletters}•
where we have defined $\tilde{q}\equiv q/(1-q)$. The average number of stars sampled in bin $k$ will therefore be
\begin{mathletters}
\begin{eqnarray}
\langle r_k \rangle &=& \sum_{r=1}^{N_k} P_{k,r} r\\
&=& (1-q)^{N_k} \sum_{r=1}^{N_k} \binom{N_k}{r}\tilde{q}\,{}^r r\\
&=& (1-q)^{N_k} \;\tilde{q}\frac{d}{d\tilde{q}} \;\sum_{r=1}^{N_k} \binom{N_k}{r}\tilde{q}\,{}^r\\
&=& (1-q)^{N_k}\; \tilde{q}\frac{d}{d\tilde{q}}\; (1+\tilde{q})^{N_k}\\
&=& (1+\tilde{q})^{-N_k}\; \tilde{q}\frac{d}{d\tilde{q}}\; (1+\tilde{q})^{N_k}\\
&=& (1+\tilde{q})^{-N_k}\;\tilde{q} N_k\,(1+\tilde{q})^{N_k-1}\\
&=&  \frac{\tilde{q}}{1+\tilde{q}}\, N_k\\
&=& q\, N_k 
\end{eqnarray}•
\end{mathletters}•
which just says that if we sample a fraction $q$ of the total number of stars we expect to sample that same fraction $q$ of the stars in each bin. We can similarly calculate
\begin{mathletters}
\begin{eqnarray}
\langle r_k^2 \rangle &=&(1+\tilde{q})^{-N_k}\;\left(\tilde{q}\,\frac{d}{d\tilde{q}}\right)^2(1+\tilde{q})^{N_k-1}\\
&=&(1+\tilde{q})^{-N_k}\;\tilde{q}\,\frac{d}{d\tilde{q}}\;\tilde{q} N_k(1+\tilde{q})^{N_k-1}\\
&=&q N_k + q^2  N_k(N_k-1)
\end{eqnarray}•
\end{mathletters}•
which gives
\begin{mathletters}
\begin{eqnarray}
\Delta r_k^2 &\equiv& \langle r_k^2\rangle - \langle r_k \rangle^2= q(1-q) N_k.
\end{eqnarray}•
\end{mathletters}•
Since we expect $\Delta^{(q)}\propto \Delta r$, we thus find that 
\begin{eqnarray}
\Delta^{(q)}_{\rho}(z)= C(z) \sqrt{q(1-q)}
\end{eqnarray}•
for some function $C(z)$. We therefore find, in the limit $q\to 0$,
\begin{mathletters}
\begin{eqnarray}
\Delta_{\rho}(z) &=& \lim_{q\to 0} \frac{\Delta^{(q)}}{\sqrt{q}}\\
&=&\lim_{q\to0} C(z)\sqrt{1-q}\\
&=& C(z).
\end{eqnarray}•
\end{mathletters}•
We therefore have
\begin{mathletters}
\begin{eqnarray}
\Delta^{(q)}_{\rho}(z)&=& \Delta_{\rho}(z) \sqrt{q(1-q)}.
\end{eqnarray}•
\end{mathletters}•
Which we can use to solve for $\Delta_{\rho}(z)$ by numerically calculating $\Delta^{(q_0)}_{\rho}(z)$ at some reference value $q_0$:
\begin{mathletters}
\begin{eqnarray}
\Delta_{\rho}(z)=\frac{\Delta^{(q_0)}_{\rho}(z)}{\sqrt{q_0(1-q_0)}}.
\end{eqnarray}•
\end{mathletters}•
For the reference value $q_0=1/2$ (sampling half the stars), we therefore have
\begin{mathletters}
\begin{eqnarray}
\label{eq:factor2}
\Delta_{\rho}(z)=2\,\Delta^{(1/2)}_{\rho}(z).
\end{eqnarray}•
\end{mathletters}•
When simulating this sampling technique with a Hipparcos sample containing more stars than our data set, and extrapolating to $q=0$, we found that $\Delta_{\rho}(z)$ was more correctly given by $1.97\Delta^{(1/2)}_{\rho}(z)$. We are unsure of the reason for this discrepency. At any rate, the effect of using the factor of 1.97 instead of 2 is small compared to the remaining errors in the analysis.

For $f(w)$, the story is slightly different because of the normalization $\int\!dw\,f(w)=1$. Although $\Delta r_k$ (where here $k$ represents the $k^{\rm th}$ bin in $w$-space) is still given by $\sqrt{q(1-q)N_k}$, the quantity corresponding to $f(w)$ here is $N_k/\sum_{k^\prime} N_{k^\prime}$ in the parent set or $r_k/\sum_{k^\prime} r_{k^\prime}$ in the sample set. Since for a large number of bins,
\begin{eqnarray}
\sum_k r_k \simeq \sum_k \langle r_k \rangle = q \sum_{k^\prime} N_{k^\prime},
\end{eqnarray}•
we have
\begin{eqnarray}
\Delta^{(q)} f_k = \frac{\sqrt{q(1-q) N_k}}{q\sum_{k^\prime} N_{k^\prime}}.
\end{eqnarray}•
Similarly to the case of $\Delta(z)$, we expect, for small $q$, that
\begin{eqnarray}
\lim_{q\to0} \Delta^{(q)} f_k = \lim_{q\to0}\frac{\sqrt{q N_k}}{q\sum_{k^\prime} N_{k^\prime}}.
\end{eqnarray}•
We therefore find that
\begin{eqnarray}
\Delta_f (w)=\sqrt{\frac{q}{1-q}}\Delta^{(q)}_f(w).
\end{eqnarray}•
Since by Equation \ref{eq:hf2000}, $\rho_f(z) \sim \int\! f$ scales linearly with $f(w)$, we therefore have the same relationship for $\rho_f(z)$:
\begin{eqnarray}
\Delta_{\rho_f}(z)=\sqrt{\frac{q}{1-q}}\Delta^{(q)}_{\rho_f}(z)
\end{eqnarray}•
and, for $q=1/2$,
\begin{eqnarray}
\label{eq:factor1}
\Delta_{\rho_f}(z)=\Delta^{(1/2)}_{\rho_f}(z).
\end{eqnarray}•

\section{Appendix - Probability Densities}
\label{app:chi2}

To compute approximate probability densities, we can define a distance between expected and observed density
\begin{eqnarray}
\label{eq:chi2}
\chi^2_{\rm model} \equiv  \int_0^{z_{\rm max}}\!\!\frac{dz}{\Delta z}\, \left|\frac{\rho_{\rm model}(z)-{\rho}_{\rm obs}(z)}{\Delta_{\rho}(z)}\right|^2 \equiv \frac{X_{\rm model}}{\Delta z},
\end{eqnarray}
where $\chi^2$ is defined in the usual way as
\begin{eqnarray}
\chi^2 = \sum_{n=1}^k \frac{\left(x_n-\langle x_n\rangle\right)^2}{\Delta_n^2}
\end{eqnarray}
with $k$ the number of relevant degrees of freedom, and with $x_n$ representing the degrees of freedom. The probability density in model space would then be proportional to
\begin{eqnarray}
\label{eq:pdens}
p(\chi^2) \sim \exp\left( -\chi^2/2 \right).
\end{eqnarray}
However, we do not know the proportionality factor $\Delta z$, since we do not know the number of relevant degrees of freedom for $\Phi(z)$. Although we know the number of points we are using, the correlation between nearby points reduces the number of relevant degrees of freedom. 
This factor is important because when in the exponential in Equation \ref{eq:pdens}, it will affect the sharpness of the probability curve. A simple way to determine the degree of freedom length $\Delta z$ as well as the number $k$ of relevant degrees of freedom is to fit the probability distribution of $P(X/\Delta z)$ to a $\chi^2$ distribution:
\begin{eqnarray}
P(\chi^2)=\frac{1}{2^{\frac{k}{2}}\Gamma\left(\frac{k}{2}\right)}\left(\chi^2\right)^{\frac{k}{2}-1}e^{-\chi^2/2}.
\end{eqnarray}
The reason this method is only approximate is that there is degeneracy between the two parameters we are fitting, $k$ and $1/\Delta z$. To see this, note that the position of the peak of the $\chi^2$ distribution grows linearly with $k$. Clearly, the position of the peak of $P(X/\Delta z)$ also grows linearly with $1/\Delta z$. Although other aspects of $P(\chi^2)$ also depend on $k$, this degeneracy still persists and makes it difficult to fit the distributions. While it does not solve this degeneracy problem, it will prove slightly simpler to fit the cumulative distribution $C(X)=\int_0^X\!dX^\prime\, P(X^\prime)$, given by the incomplete Gamma function:
\begin{eqnarray}
C(\chi^2)=\frac{\Gamma\left(\frac{k}{2},\chi^2\right)}{\Gamma\left(\frac{k}{2}\right)},
\end{eqnarray}
Another factor that complicates this is that $\Phi(z)$ is a functional depending on all the values of $f(w)$. What this implies is that although the errors on $f(w)$ and $\rho(z)$ are approximately Gaussian, the errors on $\Phi(z)$ will not be. Figure \ref{fig:PXA} below shows the $X$ distributions for A and F stars with best fit $\Delta z$ and $k=12$ and 8 (respectively) degrees of freedom. However, because of the degeneracy between $k$ and $1/\Delta z$, the best fit values are very approximate and the derived probability density $p(\Sigma_D)$ should therefore only be regarded as qualitative. 
\begin{figure}[h]
\centering{}\caption{ Probability density for $X$ computed from A and F stars by sampling using statistical procedure defined above. Superimposed $\chi^2$ distributions with 12 and 8 relevant degrees of freedom, respectively.}
\label{fig:PXA}
\includegraphics[scale=0.6]{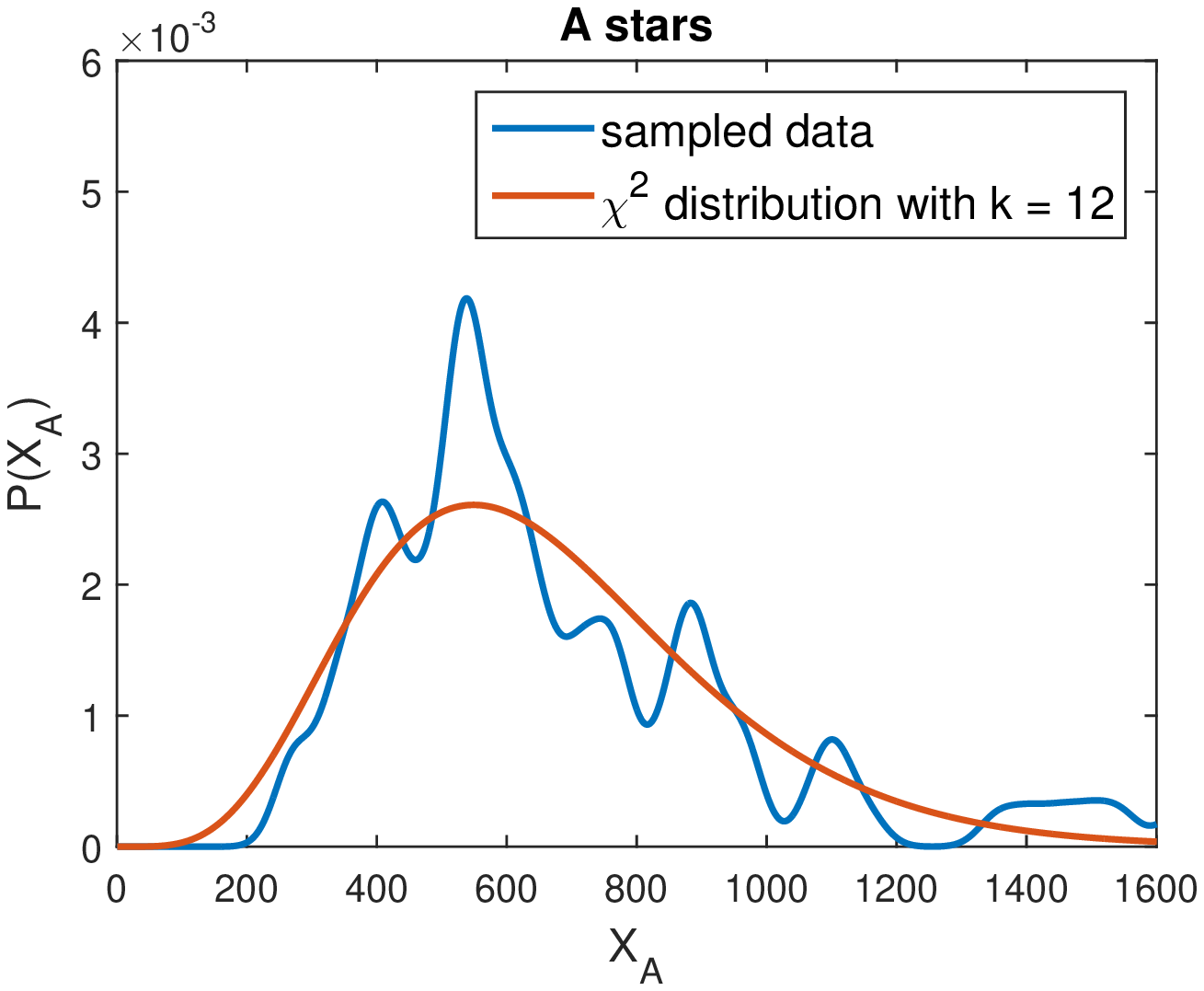}\includegraphics[scale=0.6]{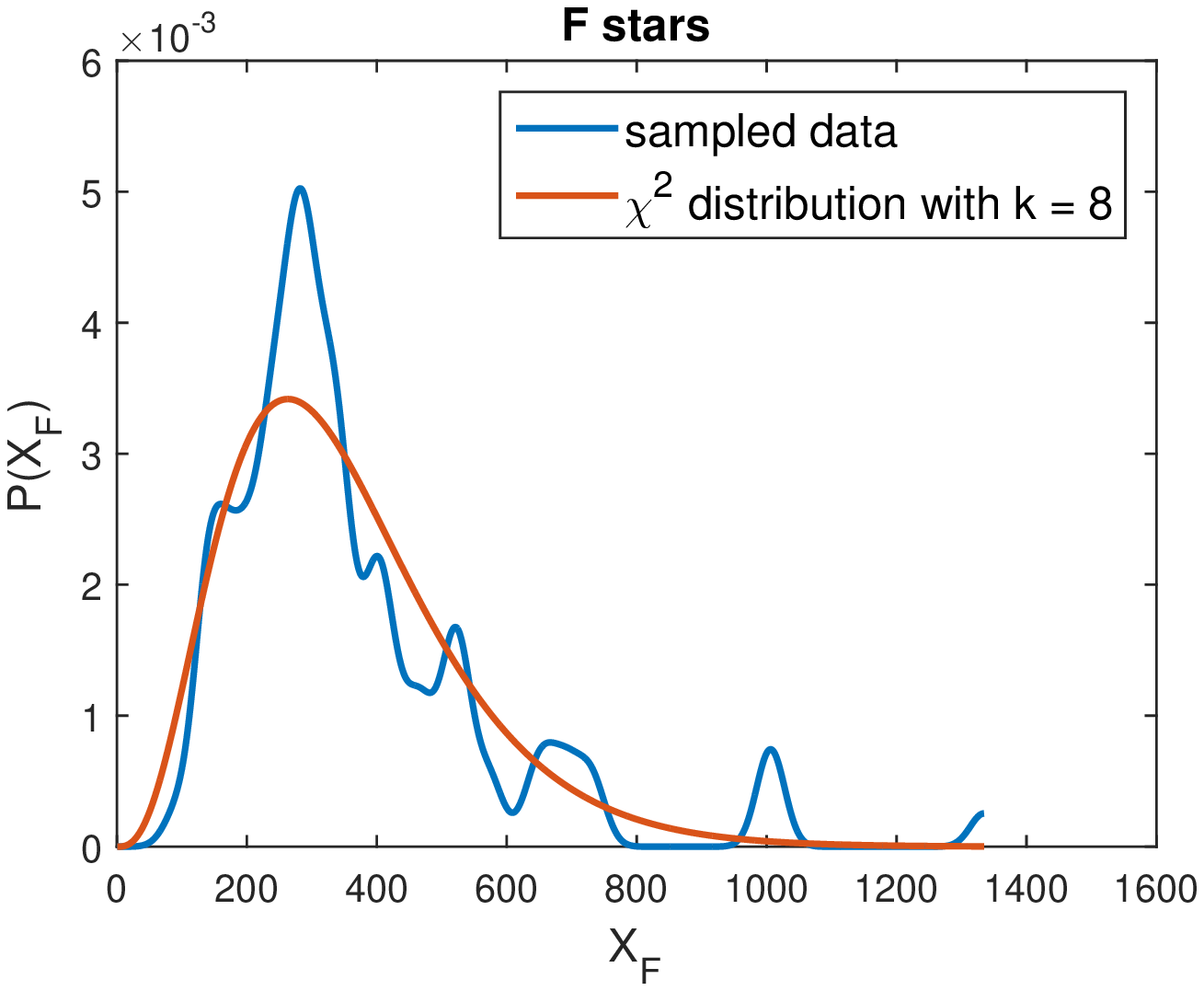}
\end{figure}

\section{Appendix - Non-Equilibrium Method}
\label{sec:noneq}

As explained in Section \ref{sec:noneqintro}, we expect that the HF relation will hold for long-time averages. We will demonstrate that
\begin{eqnarray*}
\label{eq:hfavg}
\overline{\rho(z)}_{[\Phi]}\overline{\rho(0)^{-1}}_{[\Phi]}&=&\int\!dw\;\left.\overline{{f_{z=0}}}_{[\Phi]}(\sqrt{w^2+2\Phi(z)})\right.
\end{eqnarray*}
where the $\overline{\phantom{(}\,\cdot\,\phantom{)}}_{[\Phi]}$ represents the time average under evolution in the potential $\Phi(z)$. However, if this potential $\Phi$ is the same potential as under the square root $\sqrt{w^2+\Phi(z)}$, then \ref{eq:hfavg} will be satisfied trivially for any initial conditions. This can be demonstrated by taking the time-dependent density and in-plane velocity distributions to be:
\begin{mathletters}
\begin{eqnarray}
\rho(z,t)&=&\frac{1}{A_0}\sum_i\delta(z-z_i(t))\\
f_{z=0}(w,t)&=&\frac{\sum_i\delta(w-w_i(t))\,\theta(|z_i(t)|<{\epsilon}/{2})}{\sum_j\theta(|z_j(t)|<\epsilon/{2})}
\end{eqnarray}
\end{mathletters}
where the sums are over all stars in the tracer population, $A_0$ is the cross-sectional area of our sample, and $\theta(|z_i(t)|<\epsilon/2)$ assures that $z_i(t)$ is within some appropriate small distance $\epsilon/2$ of the plane. We can now compute the long-time average:
\begin{mathletters}
\begin{eqnarray}
\overline{\rho(z)}&=&\lim_{t\to\infty}\frac1t\int_0^t\!\!dt^\prime\;\rho(z,t^\prime)\\
&=&\frac{1}{A_0}\sum_i\lim_{t\to\infty}\frac1t\int_0^t\!\!dt^\prime\;\delta(z-z_i(t^\prime))\\
&=&\frac{1}{A_0}\sum_i\lim_{N_i\to\infty}\frac{1}{N_iT_{i}/2}N_i\int_0^{T_i/2}\!\!dt\;\delta(z-z_i(t))\\
&=&\frac{1}{A_0}\sum_i\frac{2}{T_{i}}\int_{-z_{\rm max}}^{z_{\rm max}}\!\!\frac{dz_i}{w_i(z_i)}\;\delta(z-z_i)\\
&=&\frac{1}{A_0}\sum_i\frac{2}{T_{i}|w_i(z)|}
\label{eq:qed}
\end{eqnarray}
\end{mathletters}
where we have used the fact that the trajectories of the stars are periodic with individual periods $T_i$, and where $|w_i(z_i)|$ is star $i$'s speed (fixed by energy conservation) at height $z_i$. We can proceed similarly for the velocity distribution. For a large number of stars, we can write the denominator as 
\begin{eqnarray}
\sum_i\theta(|z_i(t)|<\epsilon/2)=\epsilon\, \rho(0,t)\,  A_0.
\end{eqnarray}
We now proceed to write
\begin{mathletters}
\begin{eqnarray}
\overline{f_{z=0}(w)}&=&\lim_{t\to\infty}\frac1t\int_0^t\!\!dt^\prime\;f_{z=0}(w,t^\prime)\\
&=&\sum_i\lim_{t\to\infty}\frac1t\int_0^t\!\!dt^\prime\;\frac{\delta(w-w_i(t^\prime))\,\theta(|z_i(t^\prime)|<\epsilon/2)}{\epsilon\, \rho(0,t)\,  A_0}
\end{eqnarray}
\end{mathletters}
Now, when star $i$ is at $z=0$, its velocity will be maximum and will be equal to $w_i=\sqrt{2E_i}$. The amount of time that star $i$ spends between $z=\mp\epsilon/2$ will therefore be given by $\epsilon/\sqrt{2E_i}$. In each period, the integral will therefore receive a contribution $\epsilon/\sqrt{2E_i}\;\delta(w-\sqrt{2E_i})$ on the way up and $\epsilon/\sqrt{2E_i}\;\delta(w+\sqrt{2E_i})$ on the way down. We can achieve this by replacing
\begin{mathletters}
\begin{eqnarray}
\theta(|z_i(t^\prime)|<\epsilon)&=&\frac{\epsilon}{\sqrt{2E_i}}\sum_{n_i=0}^{N_i}\delta(t-t_{0i}-n_iT_i/2)\\
&=&\frac{\epsilon}{\sqrt{2E_i}}\sum_{n_i=0}^{N_i/2}\left(\delta(t-t_{0i}-n_iT_i)+\delta(t-t_{0i}-(n_i+1/2)T_i)\right)
\end{eqnarray}
\end{mathletters}
where the first crossing of the plane for star $i$ happens at time $t_{0i}$. We thus have
\begin{mathletters}
\begin{eqnarray}
\overline{f_{z=0}(w)}&=&\frac{1}{\epsilon A_0} \sum_i\frac{\epsilon}{\sqrt{2E_i}}\lim_{N_i\to\infty}\frac{1}{N_iT_i/2}\int_0^{N_iT_i/2}\!\!dt\;\frac{1}{\rho(0,t)}...\\
&&\quad \times\left[\delta(w-{\rm sign}(w_i(t_{0i}))\sqrt{2E_i})\,\sum_{n_i=0}^{N_i/2}\delta(t-t_{0i}-n_iT_i)\right. ...\\
&&\quad+\left.\delta(w+{\rm sign}(w_i(t_{0i}))\sqrt{2E_i})\,\sum_{n_i=0}^{N_i/2}\delta(t-t_{0i}-(n_i+1/2)T_i)\right]\\
&=&\frac{1}{A_0} \sum_i\frac{1}{\sqrt{2E_i}}\lim_{N_i\to\infty}\frac{1}{N_iT_i/2}...\\
&&\quad \times\left[\delta(w-{\rm sign}(w_i(t_{0i}))\sqrt{2E_i})\,\sum_{n_i=0}^{N_i/2}\frac{1}{\rho(0,t_{0i}+n_iT_i)}\right. ...\\
&&\quad+\left.\delta(w+{\rm sign}(w_i(t_{0i}))\sqrt{2E_i})\,\sum_{n_i=0}^{N_i/2}\frac{1}{\rho(0,t_{0i}+(n_i+1/2)T_i)}\right].
\end{eqnarray}
\end{mathletters}
Now, unless the periods of $\rho(0,t)$ and of the trajectories of the individual stars divide each other, $\sum_{n_i=0}^{N_i/2}{1}/{\rho(0,t_{0i}+n_iT_i)}$ will just be, in the large $N_i$ limit, $N_i/2$ times a time average of $1/\rho(0,t)$. Since the stars with periods dividing that of $\rho(0,t)$ is a set of measure zero, we have
\begin{mathletters}
\begin{eqnarray}
&=&\frac{1}{A_0} \sum_i\frac{1}{\sqrt{2E_i}}\lim_{N_i\to\infty}\frac{1}{N_iT_i/2}...\\
&&\quad \times\left[\delta(w-{\rm sign}(w_i(t_{0i}))\sqrt{2E_i})\,\frac{N_i}{2}\,\overline{\rho(0)^{-1}}\right. ...\\
&&\quad+\left.\delta(w+{\rm sign}(w_i(t_{0i}))\sqrt{2E_i})\,\frac{N_i}{2}\,\overline{\rho(0)^{-1}}\right]\\
&=&\frac{1}{A_0} \sum_i\frac{1}{\sqrt{2E_i}}\frac{1}{T_i}\,\overline{\rho(0)^{-1}}\,\left[\delta(w-\sqrt{2E_i})+\delta(w+\sqrt{2E_i})\right]\\
&=&\frac{1}{A_0} \sum_i\frac{1}{\sqrt{2E_i}}\frac{1}{T_i}\,\overline{\rho(0)^{-1}}\,2\sqrt{2E_i}\,\delta(w^2-2E_i)\\
&=&\frac{\overline{\rho(0)^{-1}}}{A_0} \sum_i\frac{2}{T_i}\,\delta(w^2-2E_i).
\end{eqnarray}
\end{mathletters}
We can now evaluate $\left.\overline{{f_{z=0}}(w^\prime)}\right|_{w^\prime=\sqrt{w^2+2\Phi(z)}}$, using the fact that $w_i(z)^2/2+\Phi(z)=E_i$:
\begin{mathletters}
\begin{eqnarray}
\left.\overline{{f_{z=0}}(w^\prime)}\right|_{w^\prime=\sqrt{w^2+2\Phi(z)}}&=&\frac{\overline{\rho(0)^{-1}}}{A_0} \sum_i\frac{2}{T_i}\,\delta(w^2-2(E_i-\Phi(z)))\\
&=&\frac{\overline{\rho(0)^{-1}}}{A_0} \sum_i\frac{2}{T_i}\,\delta(w^2-w_i(z)^2)
\end{eqnarray}
\end{mathletters}
where $|w_i(z)|$ is the speed of star $i$ at height $z$. This gives
\begin{mathletters}
\begin{eqnarray}
\int\!dw\;\left.\overline{f_{z=0}(w^\prime)}\right|_{w^\prime=\sqrt{w^2+2\Phi(z)}}&=&\int\!dw\;\frac{\overline{\rho(0)^{-1}}}{A_0} \sum_i\frac{2}{T_i}\,\delta(w^2-w_i(z)^2)\\
&=&\frac{\overline{\rho(0)^{-1}}}{A_0} \sum_i\frac{2}{T_i|w_i(z)|}
\end{eqnarray}
\end{mathletters}
which, by Equation \ref{eq:qed}, is equal to $\overline{\rho(z)}$.\\

Q.E.D.

\end{document}